\documentclass{aa}
\usepackage[varg]{txfonts}
\usepackage{graphicx}
\usepackage{txfonts}
\usepackage{natbib}
\usepackage{verbatim}
\usepackage[pdftex]{color}
\usepackage[dvipsnames]{xcolor}
\usepackage{hyperref}
\usepackage{ulem}
\usepackage{subfigure}
\hypersetup{
    colorlinks = true,
    citecolor ={blue}
}
\usepackage{mathrsfs}
\bibpunct{(}{)}{;}{a}{}{,}

\newcommand{\G}[1]{$Gaia$}

\usepackage{booktabs}
\definecolor{ultramarine}{rgb}{0.07, 0.1, 0.6} 
\definecolor{myblue}{rgb}{0.07, 0.2, 0.6} 
\definecolor{dopal}{rgb}{.70, .25, .05}

\begin{document}

\title{A magnitude-limited catalogue of unresolved white dwarf-main sequence binaries from \G\, DR3 
}

\author{Alberto Rebassa-Mansergas\inst{1,2}\thanks{E-mail: alberto.rebassa@upc.edu},
Enrique Solano\inst{3}, Alex J. Brown\inst{4}, Steven G. Parsons\inst{5}, Raquel Murillo-Ojeda\inst{3}, Roberto Raddi\inst{1}, Maria Camisassa\inst{1}, Santiago Torres\inst{1,2}, Jan van Roestel\inst{6}}
\institute{Departament de F\'isica, 
           Universitat Polit\`ecnica de Catalunya, 
           c/Esteve Terrades 5, 
           08860 Castelldefels, 
           Spain
           \and
            Institut d'Estudis Espacials de Catalunya (IEEC), C/Esteve Terradas, 1, Edifici RDIT, 08860 Castelldefels, Spain
           \and
           Centro de Astrobiología (CAB), CSIC-INTA, Camino Bajo del Castillo s/n, 28692 Villanueva de la Cañada, Madrid, Spain
           \and
           Hamburger Sternwarte, University of Hamburg, Gojenbergsweg 112, 21029 Hamburg, Germany
           \and
           Astrophysics Research Cluster, School of Mathematical and Physical Sciences, University of Sheffield, Sheffield S3 7RH, UK
           \and        
           Anton Pannekoek Institute for Astronomy, University of Amsterdam, 1090 GE, Amsterdam, The Netherlands
           }

\date{Received ; accepted }

\abstract{
Binary stars containing  a white dwarf and a  main-sequence star, WDMS
binaries,  can be  used to  study a  wide range of  aspects of  stellar
astrophysics.  }
{
We build a magnitude-limited sample of unresolved WDMS binaries from
\G\, DR3 to enlarge these studies.  } 
{
We look for WDMS with available spectra whose location in the \G\,
colour-magnitude diagram bridges between the evolutionary sequences of
single white dwarfs and the main-sequence. To exclude spurious sources
we apply quality cuts on the \G\, photometry and astrometry and we fit
the SED (spectral energy distribution) of the objects with VOSA
(Virtual Observatory SED Analyser) to exclude single sources. We
further clean the sample via visual inspection of the \G\, spectra and
publicly available images of the objects. We re-fit the SEDs of the
finally selected WDMS with VOSA using composite models to measure
their stellar parameters and we search for eclipsing systems by
inspecting available ZTF and CRTS light curves.  }        
{
The catalogue  consists of 1312 WDMS  and we manage to  derive stellar
parameters for  435. This is  because most  WDMS are dominated  by the
main-sequence companions, making it  hard to derive parameters for the
white dwarfs.  We also  identify 67 eclipsing  systems and  estimate a
lower  limit to  the completeness  of  the sample  to be  $\simeq$50\%
($\simeq$5\% if  we consider that not  all WDMS in the  studied region
have \G\, spectra).
}
{
Our catalogue increases by one order of magnitude the volume-limited
sample we presented in our previous work. Despite the fact that the
sample is incomplete and suffers from heavy observational biases, it
is well characterised and can therefore be used to further constrain
binary evolution by comparing the observed properties to those from
synthetic samples obtained modeling the WDMS population in the
Galaxy, taking into account all selection effects.  }

\keywords{(Stars:) white dwarfs; (Stars:) binaries (including multiple): close}
\titlerunning{WDMS binaries from \G\, DR3}
\authorrunning{Rebassa-Mansergas et al.}

\maketitle

\section{Introduction}
\label{introduction}

White dwarf-main sequence (WDMS) binaries are binary stars formed by a
white  dwarf, the  most common  stellar remnant,  and a  main-sequence
star. They descend  from main-sequence binaries in  which the primary,
more massive  star, had time to  evolve out of the  main-sequence. Two
general pathways lead to the formation of a WDMS.

The first  one involves mass  transfer interactions that  usually take
place once  the primary becomes  a red  giant, or an  asymptotic giant
star. That is, the initial  main-sequence binary orbital separation is
short enough ($\lesssim 10$ AU; \citealt{farihi10}) for the giant star
to overfill its Roche-lobe and to transfer mass to the secondary, less
massive,  companion.  Given  that   the  mass  transfer  is  generally
dynamically unstable, the system is thought to evolve through a common
envelope phase  \citep{Paczynski76, Webbink2008} in which  the core of
the giant  and the  secondary star are  surrounded by  common material
formed by the outer layers of  the giant -- that have been transferred
to  but not  accreted by  the companion  -- and  friction considerably
reduces  the  orbital  separation,  hence  orbital  period  to  a  few
hours/days  \citep{Rebassa-Mansergasetal08,  Nebotetal11}.  WDMS  with
orbital periods as large as $\simeq$1000 days are also suggested to be
the  outcome  of  common envelope  evolution  \citep{Yamaguchietal24},
being   stable   non-conservative   mass  transfer   the   alternative
evolutionary path for such long-period systems \citep{Hallakounetal24,
  Garbuttetal24}.  It  is  expected that  these  post-common  envelope
binaries account for  approximately 25\% of the  initial main-sequence
binaries \citep{Willems+Kolb04}.

The second  scenario, encompassing  the remaining $\simeq$75\%  of the
cases, does not involve mass transfer episodes, since the initial main
sequence binary orbits are wide enough to avoid them, and consequently
the  primary star  evolves like  a single  star. In  these cases,  the
orbital periods of  the WDMS binaries are of the  order of hundreds to
thousands of days.

Both  wide  WDMS  binaries  that   evolved  like  isolated  stars  and
post-common envelope binaries  have been of extreme value  to tackle a
wide diversity of issues. For instance, since the white dwarfs in wide
WDMS binaries  can be used  to measure  stellar ages, studies  of such
systems    have     constrained    the     age-metallicity    relation
\citep{rebassa-mansergasetal16-2,  rebassa-mansergasetal21-2} and  the
age-velocity  dispersion  relation   \citep{Raddi2022}  in  the  solar
neighbourhood,  as  well  as  the  age-activity-rotation  relation  of
low-mass    main-sequence   stars    \citep{rebassa-mansergasetal13-1,
  rebassa-mansergasetal23,  Chittietal24}. They  can also  be used  to
test  the  white  dwarf  mass-radius  relation  \citep{Arseneauetal24,
  Raddietal25}    and     the    initial-to-final     mass    relation
\citep{Zhaoetal12,   Barrientos2021}.   On   the  other   hand,  close
post-common  envelope binaries  allow constraining  the efficiency  of
common    envelope    ejection   \citep{Zorotovic2010,    Camacho2014,
  Cojocaru2017,  Grondinetal24},  the  mass-radius relation  of  white
dwarfs     \citep{Parsons2017},    low-mass     main-sequence    stars
\citep{Parsons2018}  and even  brown dwarfs  \citep{Parsonsetal25} and
sub-dwarf  stars \citep{Rebassa2019b}  via the  analysis of  eclipsing
systems,     the      origin     of     low-mass      white     dwarfs
\citep{rebassa-mansergasetal11},  angular   momentum  losses   due  to
magnetic braking  \citep{Schreiber2010, Zorotovic2016} and  the origin
of magnetism in white dwarfs \citep{Marsh2016, Schreiber2021}.

The  first large  catalogue of  $\simeq3200$ WDMS  binaries was  built
thanks   to   the   mining   of   the   Sloan   Digital   Sky   Survey
\citep[SDSS;][]{Eisensteinetal11}        spectroscopic        database
\citep{rebassa-mansergasetal12-1,  rebassa-mansergasetal16-1}, closely
followed  by the  spectroscopic  catalogue  of $\simeq$900  additional
systems \citep{Renetal18}  from the Large Sky  Area Multi-Object Fiber
Spectroscopic  Telescope  \citep[LAMOST;][]{Cui2012RAA}.  Since  these
samples are  largely affected by  selection effects, in  particular by
the fact that earlier type than M companions outshine the white dwarfs
in  the optical  \citep{rebassa-mansergasetal10-1}, efforts  have been
placed  to  combine  ultraviolet photometry  with  optical  photometry
and/or spectroscopy  that allowed  the identification of  thousands of
WDMS binaries  containing F, G and  K companions \citep{Parsonsetal16,
  Rebassa2017,      Renetal20,       Anguianoetal22,      Nayaketal24,
  Sidharthetal24}.

A  potential  issue from  the  above  studies  is  that they  are  all
magnitude-limited, which  makes it difficult to  unveil the underlying
population unless population synthesis  studies are taken into account
\citep{Davisetal10, Toonen+Nelemans13,  Torresetal22}. In  this sense,
the  astrometry   and  photometry  provided  by   the  \G\,  satellite
\citep{Gaia2018,  Gaia2023}  allowed  mitigating this  effect,  as  it
became  possible  to build  the  first  volume-limited sample  of  112
well-characterised  candidates  within  100  pc from  the  early  data
release  3 \citep{rebassa-mansergasetal21-1}.   In  this analysis,  we
defined    a    region    in    the    \G\,    $G_\mathrm{abs}$    vs.
$G_\mathrm{BP}-G_\mathrm{RP}$ diagram  to exclude single  white dwarfs
and  main-sequence stars  and  derived white  dwarf and  main-sequence
stellar  parameter  distributions  that clearly  differed  from  those
obtained   from  magnitude-limited   samples.    Moreover,  a   direct
comparison with  the parameter  distributions obtained  from numerical
simulations that reproduced the \G\, population in the Galaxy provided
additional valuable  insight into binary star  formation and evolution
\citep{Santos-Garcia2025}.     Unfortunately,    despite    being    a
volume-limited sample,  the \G\, catalogue  was revealed to  be highly
incomplete, as  most of the  WDMS binaries  are expected to  have \G\,
colours    very   similar    to   those    of   main-sequence    stars
\citep{Santos-Garcia2025}, which  were excluded from the  analysis. As
mentioned above,  these systems  are difficult  to identify  since the
main-sequence companions outshine  the white dwarfs in  the optical. A
promising  way  to   move  forward  is  to  make   use  of  artificial
intelligence algorithms to  differentiate between single main-sequence
stars and  WDMS binaries  via the analysis  of available  \G\, spectra
from   its   data    release   3   \citep{Echeverryetal22,   Lietal25,
  Perez-Couto2025}.

In  this work,  our motivation  is to  build up  the \G\,  WDMS binary
sample  we  presented   in  \citet{rebassa-mansergasetal21-1}  by  not
implementing any distance cut, but maintaining the focus in the bridge
region  of  the \G\,  colour-magnitude  diagram  between single  white
dwarfs and single main-sequence stars.  That is, we aim at providing a
magnitude-limited but well-characterised sample  from \G\, that can be
directly compared to the output of numerical simulations such as those
we implemented in \citet{Santos-Garcia2025}.  This will allow deriving
further constraints on binary evolution theory.

In Section\,\ref{s-sample} we introduce the  WDMS binary sample and in
Section\,\ref{s-param}   we   attempt    to   derive   their   stellar
parameters. In  Section\,\ref{s-comp} we  compare our sample  to other
works  in  the  literature.   In  Section\,\ref{s-eclip}  we  identify
eclipsing  WDMS   among  our   objects  and   conclude  the   work  in
Section\,\ref{s-concl}.

\section{The WDMS sample}
\label{s-sample}

\begin{figure*}
    \centering
    \includegraphics[angle=-90,width=0.9\columnwidth]{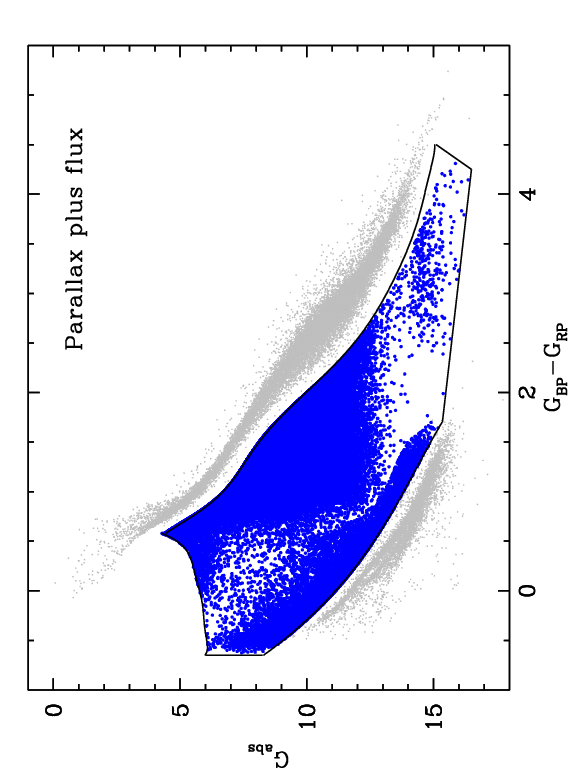}
    \includegraphics[angle=-90,width=0.9\columnwidth]{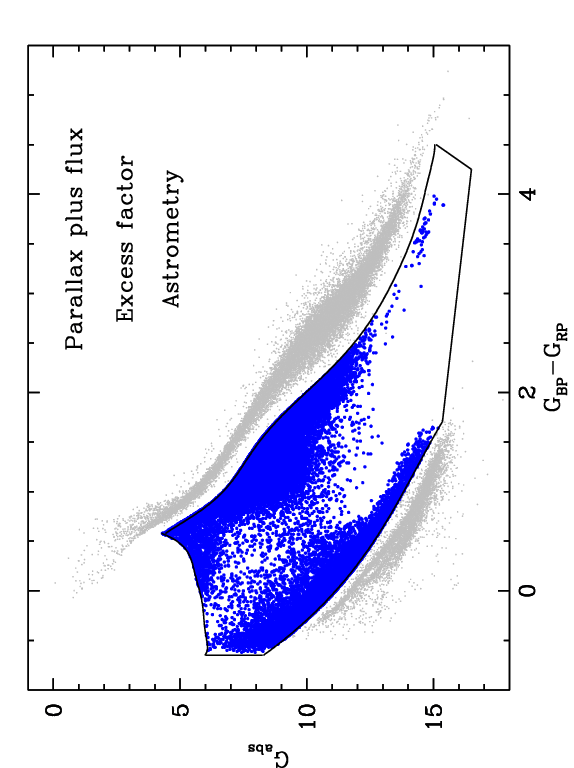}
    \includegraphics[angle=-90,width=0.9\columnwidth]{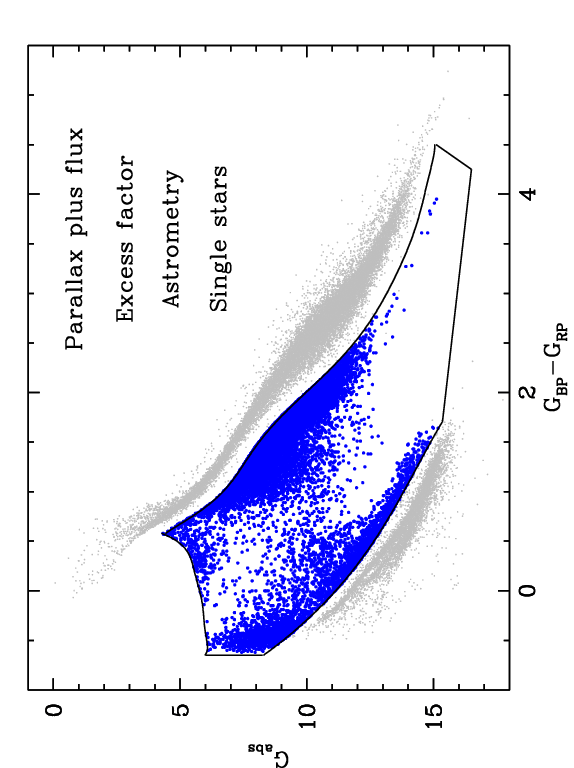}
    \includegraphics[angle=-90,width=0.9\columnwidth]{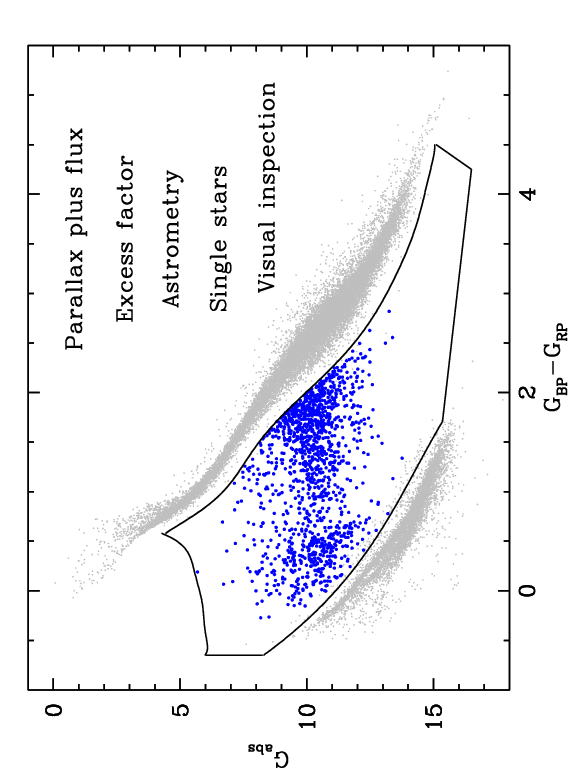}
    \caption{Results of our criteria imposed on the \G, date release 3
      data   base   to   select   WDMS  binaries   (see   details   in
      Section\,\ref{s-sample}). Top left: in blue the 126\,787 sources
      within the WDMS  binary region defined by the  black solid lines
      \citep{rebassa-mansergasetal21-1}  with  available  spectra  and
      satisfying parallax  and flux  relative errors above  10\%.  The
      gray  dots  illustrate the  expected  location  of single  white
      dwarfs  and  main-sequence  stars.  Top right:  the  same  after
      applying    our   excess    factor    and   astrometry    (RUWE,
      astrometric\_excess\_noise  and astrometric\_excess\_noise\_sig)
      cuts, which leaves 62\,386 objects.  Bottom left: the same after
      excluding   single    stars   using   VOSA,    leaving   13\,905
      candidates.  Bottom  right: the  final  catalogue  of 1312  WDMS
      binaries after visual inspection of the \G\, spectra.}
    \label{fig_sel}
\end{figure*}

We      follow      basically      the      same      criteria      as
\citet{rebassa-mansergasetal21-1} to build the WDMS binary sample from
\G, data  release (DR)  3 \citep{Gaia2023}. That  is, we  consider all
objects in  the intermediate  region between  single white  dwarfs and
single main-sequence  stars in the \G\,  colour-magnitude diagram with
parallax  over error  and $G$,  $G_\mathrm{BP}$, $G_\mathrm{RP}$  flux
over error parameters greater than 10. However, in this work we do not
impose any  distance limit  and we  focus the  search to  objects with
available \G\, spectra.  As we will see later,  this condition (having
available   \G\,   spectra)   affects    the   completeness   of   our
catalogue.  However, it  is required  since we  intend to  measure the
stellar   parameters  of   the   identified   candidates  from   these
spectra. Moreover,  visual inspection greatly helps  in confirming the
binary  nature  of  the  candidates   by  identification  of  the  two
components  in  the  available  spectra.  This  resulted  in  126\,787
selected   sources,   illustrated   in   the   top   left   panel   of
Figure\,\ref{fig_sel}.

To reduce contaminants, defined as sources that are not WDMS, from our
selected candidates, we first applied a condition on the excess factor
parameter  provided  by  \G..  This   parameter  is  defined  as  $C=(
F_\mathrm{BP}+F_\mathrm{RP})/F_\mathrm{G}$  \citep{Evansetal18}, where
$F$ denotes flux in the \G\, bands, and it should be close to 1. Thus,
any object  with deviations from  $C=1$ may be associated  to internal
calibration issues. It  has to be emphasised that one  of the possible
causes for having a large excess factor is binarity, that is \G\, sees
two  sources that  form  a  partially resolved  system.   As noted  by
\citet{Rielloetal21},  $C$ is  colour  dependent and  so they  defined
$C^{*}  = C-f(G_\mathrm{BP},G_\mathrm{RP})$,  where  the function  $f$
provides the  expected excess at  a given  colour. In other  words, by
correcting  the expected  excess, any  object with  a $|C^{*}|$  value
larger than zero  can be considered as a potential  source affected by
calibration issues.  It depends on  the user  to be as  restrictive as
necessary to  exclude such sources. To  that end we followed  the same
approach as \citet{rebassa-mansergasetal21-1}  and removed all sources
with:

\begin{eqnarray}
      |C^{*}|\geq0.3, &  G_\mathrm{BP}-G_\mathrm{RP}<0.5;\\
  |C^{*}|\geq0.2, &  0.5\leq G_\mathrm{BP}-G_\mathrm{RP}\leq4;\\
 |C^{*}|\geq0.1,  &  G_\mathrm{BP}-G_\mathrm{RP}>4.
\end{eqnarray}

This cut  reduced the WDMS candidates  to 82\,021. It should  be noted
that these cuts are different  from the traditional criteria suggested
by  \citet{Rielloetal21}, which  is based  not only  on the  $|C^{*}|$
value but also  on its deviation $\sigma$. Thus, the  user may exclude
objects by simply  applying a $|C^{*}|>N\sigma$ cut by  fixing a value
of $N$.  However, even  when using  $N=5$, which is  expected to  be a
largely conservative  cut (meaning that  most of the  excluded sources
should indeed be  associated to spurious data), we  ended up excluding
clear WDMS  binaries from the sample.  In fact, any cut  in the excess
noise  will   unavoidably  exclude  real  WDMS,   thus  affecting  the
completeness of the  final sample. This issue will  be later discussed
in Section\,\ref{s-comp}.

So far,  we have only applied  one constraint to the  \G\, astrometry,
excluding   objects  with   parallax  relative   errors  larger   than
10\%. Thus, an additional way to further exclude contaminants from our
list  is   by  removing   objects  associated  with   bad  astrometric
solutions. In  particular, \G, provides  three parameters that  can be
used for such purpose \citep{Lindegrenetal2012}: the renormalised unit
weight error RUWE, which should be near 1.0 for point sources that are
well fitted by a single-star  model to their astrometric observations;
the   astrometric\_excess\_noise  parameter,   which  quantifies   the
agreement between the observations of  a given object and the best-fit
astrometric model; and  the astrometric\_excess\_noise\_sig parameter,
which gives the significance  of the astrometric\_excess\_noise. It is
suggested    by    \G,   that    objects    with    RUWE   $>$    1.4,
astrometric\_excess\_noise  $>2$  or  astrometric\_excess\_noise\_sig$
>2$ may  have issues  with their  astrometric solutions.  Since larger
than canonical  values of RUWE are  also possible due to  binarity, we
adopted all  sources from our  list with  RUWE values smaller  than 3,
instead of  1.4. This is  justified by looking  at figure 4  (top left
panel)  of \citet{Belokurovetal20},  where less  than 5\%  of the  801
spectroscopic binaries considered  have RUWE values larger  than 3. In
the same way, to avoid  missing possible binaries, we excluded objects
satisfying       astrometric\_excess\_noise       $>$       3       \&
astrometric\_excess\_noise\_sig $>$ 3, rather than the canonical value
of 2. As  a consequence, the number  of WDMS binaries in  our list was
further   reduced   to  62\,386   (see   the   top  right   panel   of
Figure\,\ref{fig_sel}). In  more than 99\%  of the cases,  the sources
were excluded  due to the RUWE  condition, whilst the rest  of objects
had  a   RUWE  value  smaller  than   3  but  large  values   of  both
astrometric\_excess\_noise  and astrometric\_excess\_noise\_sig.  This
means  that  our cuts  in  these  last  two parameters  are  basically
irrelevant. In Section\,\ref{s-comp} we will discuss the impact of the
adopted RUWE cut in the completeness of our catalogue.

It is  worth noting that there  are more \G, parameters  that the user
can  potentially explore  to  exclude possible  contaminants, such  as
astrometric\_sigma5d\_max\footnote{A  five-dimensional  equivalent  to
the  semi-major axis  of the  \G\, position  error ellipse.  Useful for
filtering out cases  where one of the five parameters,  or some linear
combination   of  several   parameters,  is   bad.}  \citep[see,   for
  example][]{gentile-fusilloetal19},    the    astrometric    fidelity
parameter able  to classify spurious data  \citep{Rybizkietal2022}, or
even  alternative  quantities such  as  the  local unit  weight  error
defined by \citet{Penoyreetal22}. However, we do not implement further
quality  cuts in  astrometry and  proceed  in reducing  the number  of
contaminants (in particular single sources  expected near the locus of
single white dwarfs and main-sequence stars) as follows.

\begin{figure}
    \centering
    \includegraphics[width=0.9\columnwidth]{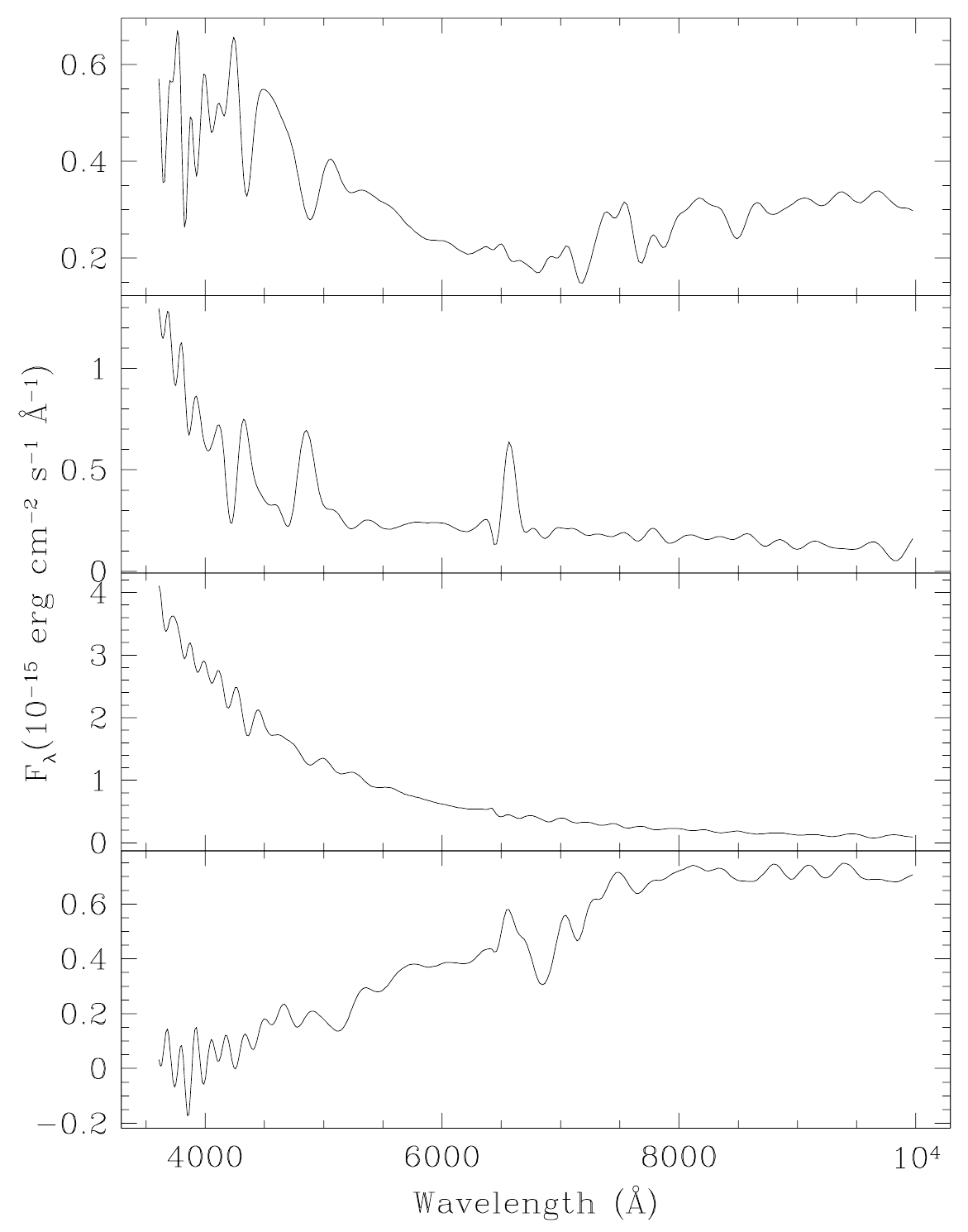}
    \caption{Example  spectra   of  a   WDMS  binary  (top,   \G\,  ID
      1057463111970047488; note that in this work we use the DR3 IDs),
      a     cataclysmic    variable     (middle    top,     \G\,    ID
      3703726255561754880),  a hot  white dwarf  (middle bottom,  \G\,
      ID1060659289192635904) and  a low-mass  low-metallicity subdwarf
      as revealed by the broad absorption feature at $\sim7000$\,\AA\,
      (bottom, \G\,  ID 1048217078174314496)  arising from  the visual
      inspection of the \G\, spectra.}
    \label{fig_spec}
\end{figure}

We  used GaiaXPy  to  convert the  \G\, spectra  of  each source  into
synthetic J-PAS \citep[Javalambre Physics of the Accelerating Universe
  Astrophysical  Survey;][]{Main-Franch2012, Benitez2014}  photometry,
which  consists  of  57  filters continuously  sampling  the  spectrum
between  3700  and  9200\AA,  thus  obtaining  their  spectral  energy
distributions  (SEDs). We  then fitted  the resulting  SEDs using  the
Virtual         Observatory         SED        Analyser         (VOSA;
\citealt{Bayoetal2008})\footnote{\url{http://svo2.cab.inta-csic.es/theory/vosa}}
tool. In the fitting process we only took into account those points in
the SED with  relative flux errors less than 10\%,  and all data above
4000\,\AA,  since  the  signal-to-noise  ratio  under  this  value  is
generally substantial. This implied  1004 objects were excluded simply
because  they   did  not   have  enough   reliable  points   in  their
SEDs. Moreover,  we adopted  the geometric  distances for  each source
from  \citet{bailerjonesetal23}, the  extinction from  the 3D  maps of
\citet{Lallement2014} and we did not use  upper limits in the fits. In
a first step, we used  the CIFIST \citep{Allardetal13} grid (effective
temperatures between  2200 and 7000\,K, surface  gravities between 4.5
and  5.5  dex,  typical  values for  main-sequence  stars,  and  solar
metallicity) to exclude 37\,712 objects  with $\chi^2$ fit values less
than 10.  In 99.5\% of  these cases, the corresponding visual goodness
of       fit       $V_\mathrm{gfb}$\footnote{See      details       at
\url{http://svo2.cab.inta-csic.es/theory/vosa/helpw4.php?otype=star&action=help&what=fit}}
was less than  2, with a maximum value of  7.2. Since $V_\mathrm{gfb}$
values of less  than 15 are usually  taken as a validation  for a good
fit, these excluded  objects should indeed be very  likely single main
sequence-stars, possibly  affected by extinction.  It  is worth noting
that  we initially  applied a  $V_\mathrm{gfb}<15$ cut  to filter  out
single  main-sequence  star candidates;  however  this  resulted in  a
non-negligible  fraction   of  excluded   WDMS  binaries  and,   as  a
consequence, we  opted for the  approach described above. In  a second
step,   we   fitted   the   remaining   24\,674   sources   with   the
\citet{Koester2010}   model  grid   of   hydrogen-rich  white   dwarfs
(effective  temperatures  between  5000  and  40\,000\,K  and  surface
gravities between 6.5 and 9.5  dex). We excluded 10\,769 sources, very
likely  single or  double\footnote{Double white  dwarfs are  generally
located   above   the  single   white   dwarf   locus  in   the   \G\,
colour-magnitude diagram, well  within our area of  study. Since their
SEDs are virtually  identical to those of single  white dwarfs, double
white dwarfs are also excluded  in this exercise.}  white dwarfs, with
$\chi^2$ fit values less than 10, which correspond to $V_\mathrm{gfb}$
values of less  than 3 in 98\%  of the cases, with a  maximum value of
8.7.  After this exercise, we were  left then with 13\,905 WDMS binary
candidates    in    our   list    (see    bottom    left   panel    of
Figure\,\ref{fig_sel}).

\begin{figure*}
    \centering
    \includegraphics[width=0.9\columnwidth]{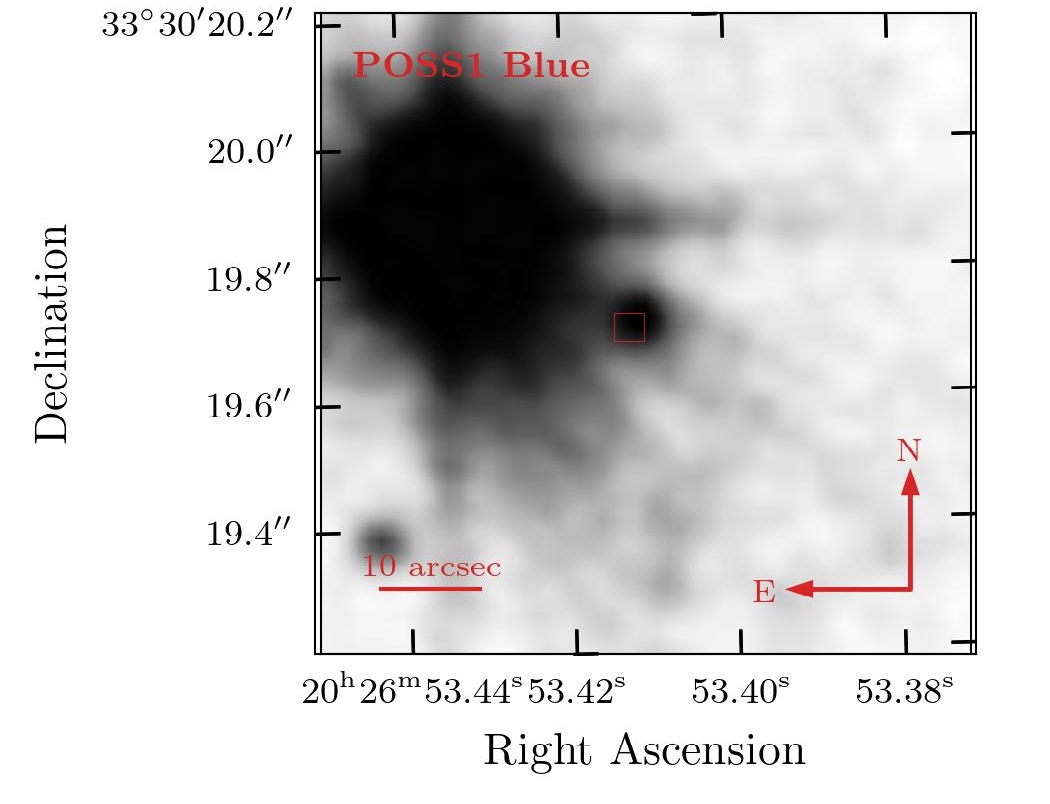}
    \includegraphics[width=0.9\columnwidth]{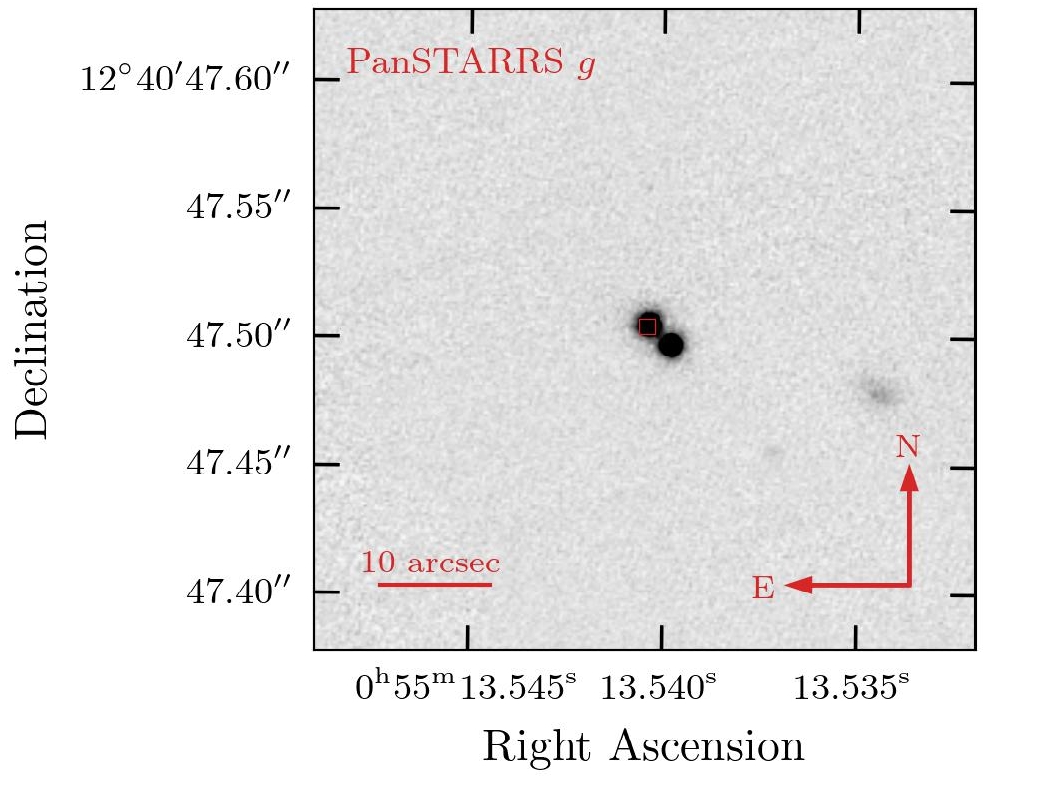}
    \caption{Example images illustrating WDMS candidates (red squares)
      for being contaminated by the  presence of nearby sources.  Left
      panel:  POSS/DSS  blue  image of  \G\,  ID  2055951194776633600.
      Right   panel:    Pan-STARSS   g-band    image   of    \G\,   ID
      2776554794743195648.}
    \label{f-cont}
\end{figure*}

We  proceeded to  visually inspect  the  \G\, spectra  of the  13\,905
candidates, which resulted in the  identification of 1312 genuine WDMS
binaries, 155 cataclysmic variables (we identify as such those objects
with  spectra displaying  prominent  and broad  Balmer emission  lines
arising from the  accretion disk) and 12\,438 other  sources.  Most of
these  other  sources  were  single  hot  white  dwarfs  and  low-mass
low-metallicity sub-dwarfs according to their \G\, spectra. Given that
the white dwarf \citet{Koester2010}  grid and the \citet{Allardetal13}
CIFIST grid  in VOSA do not  include white dwarf model  spectra hotter
than 40\,000\,K  and low metallicities stars  (only solar abundances),
respectively,  it  is  not  surprising that  these  objects  were  not
previously  considered  as  single   stars,  and  were  therefore  not
excluded.  It is  also worth noting that the  visual classification of
WDMS relies on the identification of  both stars in the spectra, which
is challenging  if one of the  components dominates the SED.   This is
worsened  due  to the  low  resolution  of  the  \G\, spectra.   As  a
consequence, this process is biased against the identification of WDMS
with mild  blue/red excess in their  spectra. In Section\,\ref{s-comp}
we  will evaluate  how these  issues  affect the  completeness of  our
sample.  Example spectra  of a WDMS binary, a  cataclysmic variable, a
single hot white dwarf and  a single low-metallicity subdwarf star can
be seen in Figure\,\ref{fig_spec}.  We  note that the above identified
CVs are not  included in our WDMS catalogue, since  we are focusing on
non-accreting binaries.  However, it is  of course plausible that some
WDMS in our sample are detached CVs crossing the period gap.

\begin{table}
\caption{WDMS selection process  from our initial sample  to the final
  catalogue.}
        \label{tab:number}
        \begin{center}
        \begin{tabular}{cccc}
            \hline\hline
            Sample & \#objects  & $f_{\rm abs}$ & $f_{\rm rel}$ \\
            \hline
        Initial                     & 126\,787 & -    & -     \\
        Excess factor cut           & 82\,021  & 0.65 & 0.65  \\
            Astrometry cuts         & 62\,386  & 0.49 & 0.76  \\
            Excluding single stars  & 13\,905  & 0.11 & 0.22  \\
            Visual inspection       & 1312     & 0.01 & 0.09  \\
           and final catalogue      &          &      &       \\
            \hline
        \end{tabular}
        \tablefoot{The  initial sample  corresponds  to those  objects
          within the  defined WDMS region  with \G\, spectra  and with
          parallax and  flux over  errors greater than  10\%.  $f_{\rm
            abs}$  and $f_{\rm  rel}$ give  the absolute  and relative
          fractions, respectively}
        \end{center}
\end{table}

To finalise, we visually inspected  the Pan-STARSS g-band and POSS/DSS
blue and red images of our objects  by eye and flagged 72 objects that
are possibly contaminated by the presence of bright nearby stars. That
is,  these  are  candidates  for  not being  real  WDMS  binaries  but
contaminated stars  by the  flux of nearby  sources. Two  examples are
shown in Figure\,\ref{f-cont}.

A  summary  of the  different  cuts  we  have applied,  including  the
fraction of excluded objects,  is provided in Table\,\ref{tab:number}.
An  excerpt of  the full  1312 WDMS  binary catalogue  is included  in
Table\,\ref{t-full}. The full table is is available at the CDS.

\begin{table*}
    \caption[]{An excerpt of the  \G\, WDMS binary catalogue.}
    \label{t-full}
    \begin{scriptsize}
\begin{center}
    \begin{tabular}{lcccccccccccc}
            \hline
        \G\, ID & Ra & Dec & par. & $G$ & $G_\mathrm{BP}$ & $G_\mathrm{RP}$  & $T_\mathrm{eff}$ (WD) & $\log g$ (WD) & M (WD) & $T_\mathrm{eff}$ (MS) & Per. & flag \\
            & (deg) & (deg)  & (mas) &(mag) & (mag) & (mag) & (K) & (dex) & M$_{\odot}$ & (K) & (days) & \\
            \hline
 1006621281985546240 &  98.37658 & 61.39074 & 13.01 & 14.30 & 15.08 & 13.36 &  14250 $\pm$ 125 & 7.90 $\pm$  0.02 &  0.55 $\pm$ 0.03 &  3400 $\pm$ 50  &       -  &  0 \\
 1019109878651542272 & 141.09262 & 51.04458 &  7.64 & 16.05 & 16.78 & 15.11 &  12000 $\pm$ 125 & 7.84 $\pm$  0.03 &  0.53 $\pm$ 0.03 &  3300 $\pm$ 50  &       -  &  0 \\
 1031612970830684800 & 123.52560 & 52.28758 &  1.94 & 17.67 & 17.71 & 17.42 &  26000 $\pm$ 500 & 7.38 $\pm$  0.04 &  0.43 $\pm$ 0.04 &  3100 $\pm$ 50  &       -  &  0 \\
 1031818515080479616 & 123.54097 & 53.32249 &  5.35 & 16.89 & 16.97 & 16.60 &  17750 $\pm$ 125 & 7.78 $\pm$  0.01 &  0.52 $\pm$ 0.02 &  2900 $\pm$ 50  &       -  &  0 \\
 1034719400416394752 & 125.26102 & 56.39990 &  6.20 & 17.34 & 17.84 & 16.55 &  11000 $\pm$ 125 & 7.92 $\pm$  0.03 &  0.55 $\pm$ 0.04 &  3100 $\pm$ 50  &       -  &  0 \\
 1041938213945851264 & 128.13788 & 61.18896 &  3.62 & 17.32 & 17.44 & 16.97 &  23000 $\pm$ 500 & 7.87 $\pm$  0.06 &  0.56 $\pm$ 0.07 &  3100 $\pm$ 50  &       -  &  0 \\
 1057463111970047488 & 172.28939 & 66.61785 &  3.41 & 17.38 & 17.71 & 16.73 &      - $\pm$   - &    - $\pm$     - &     - $\pm$    - &     - $\pm$  -  &       -  &  0 \\
 1059664849644881152 & 162.88903 & 66.56686 &  4.61 & 17.11 & 17.24 & 16.73 &  17000 $\pm$ 125 & 7.68 $\pm$  0.02 &  0.48 $\pm$ 0.02 &  3000 $\pm$ 50  &       -  &  0 \\
 1066494019444338816 & 150.70489 & 66.81527 &  4.43 & 17.09 & 17.87 & 16.09 &  12750 $\pm$ 125 & 7.82 $\pm$  0.03 &  0.52 $\pm$ 0.03 &  3200 $\pm$ 50  &       -  &  0 \\
 1068953042840614272 & 138.95581 & 67.27441 &  3.56 & 16.87 & 17.52 & 15.98 &  18250 $\pm$ 125 & 7.77 $\pm$  0.02 &  0.52 $\pm$ 0.02 &  3300 $\pm$ 50  &       -  &  0 \\
 1078143894896636672 & 157.16275 & 73.18020 & 11.34 & 16.91 & 17.93 & 15.81 &      - $\pm$   - &    - $\pm$     - &     - $\pm$    - &     - $\pm$  -  &       -  &  0 \\
 1086336218595221248 & 113.11450 & 60.69626 &  9.44 & 17.56 & 18.34 & 16.59 &      - $\pm$   - &    - $\pm$     - &     - $\pm$    - &     - $\pm$  -  &       -  &  0 \\
 1089352587012640640 & 115.49536 & 64.08135 &  5.26 & 15.71 & 15.81 & 15.37 &  20000 $\pm$ 312 & 7.23 $\pm$  0.03 &  0.36 $\pm$ 0.03 &  3100 $\pm$ 50  &       -  &  0 \\
 1091121396280101504 & 122.37888 & 62.95820 &  3.25 & 17.02 & 17.90 & 16.01 &  14750 $\pm$ 125 & 7.72 $\pm$  0.02 &  0.49 $\pm$ 0.03 &  3300 $\pm$ 50  &       -  &  0 \\
  109973634046969472 &  45.78159 & 23.29454 &  4.35 & 15.35 & 15.94 & 14.50 &      - $\pm$   - &    - $\pm$     - &     - $\pm$    - &     - $\pm$  -  &       -  &  0 \\
 1105873062754421376 &  99.53318 & 68.07844 &  3.45 & 18.16 & 18.27 & 17.82 &      - $\pm$   - &    - $\pm$     - &     - $\pm$    - &     - $\pm$  -  &       -  &  1 \\
 1106926566694585984 &  94.80641 & 69.19361 &  3.22 & 16.89 & 16.83 & 16.92 &  23000 $\pm$ 500 & 7.33 $\pm$  0.05 &  0.40 $\pm$ 0.05 &  2900 $\pm$ 50  &       -  &  0 \\
 1125637127860398848 & 131.85117 & 76.31376 &  8.38 & 16.61 & 17.49 & 15.62 &      - $\pm$   - &    - $\pm$     - &     - $\pm$    - &     - $\pm$  -  &       -  &  0 \\
 1128036811987813888 & 155.08710 & 77.28237 &  5.40 & 17.50 & 18.07 & 16.67 &  11500 $\pm$ 125 & 7.97 $\pm$  0.03 &  0.58 $\pm$ 0.04 &  3100 $\pm$ 50  & 1.38206$^{(0)}$  &  0 \\
 1132443276634291200 & 146.52269 & 80.02909 &  4.36 & 17.44 & 18.15 & 16.51 &  12250 $\pm$ 125 & 7.90 $\pm$  0.03 &  0.55 $\pm$ 0.03 &  3200 $\pm$ 50  &       -  &  0 \\
         $\ldots$  & $\ldots$ &  $\ldots$ & $\ldots$ & $\ldots$ & $\ldots$ & $\ldots$& $\ldots$& $\ldots$& $\ldots$& $\ldots$ & $\ldots$ & $\ldots$  \\
            \noalign{\smallskip}
            \hline
    \end{tabular}
\end{center}
    \end{scriptsize}
\tablefoot{Included  are   the  \G\,  source  IDs,   the  coordinates,
  parallaxes, magnitudes, stellar parameters  (note that the effective
  temperature errors  are simply half  the step between  the available
  models),  orbital periods  for the  67 identified  eclipsing systems
  (also including  a note indicating  if the  system is new  from this
  work  or has  been published  before; Section\,\ref{s-eclip})  and a
  flag  indicating whether  (1)  or  not (0)  the  object is  possibly
  contaminated by the  presence of nearby bright  sources.  We provide
  the stellar parameters only for all WDMS with effective temperatures
  higher  than  10\,000 K  and  masses  above 0.35\,M$_{\odot}$.   The
  objects are given  in order of ascendant source ID.  The Per. column
  is  associated  to the  following  references:  (0) This  work;  (1)
  \citet{Brownetal2023};      (2)     \citet{Bruch+Diaz1998};      (3)
  \citet{Chenetal2020};      (4)      \citet{Kosakowski2022};      (5)
  \citet{Mowlavietal2023};     (6)      \citet{Nebotetal2009};     (7)
  \citet{ODonoghueetal2003};     (8)      \citet{Parsons2013};     (9)
  \citet{Parsonsetal2015};  (10)   \citet{Priyatikantoetal2022};  (11)
  \citet{Pyrzasetal2009}; (12) \citet{Pyrzasetal2012}.}
\end{table*}

\section{Stellar parameters}
\label{s-param}

In this  section we attempt  to derive  the stellar parameters  of the
1312 WDMS binary candidates. To that end we first use the two-body fit
implemented  in VOSA.  In this  case, we  complemented the  \G\, J-PAS
synthetic     photometry     with    Galaxy     Evolution     Explorer
\citep[\textit{GALEX};][]{Bianchi17},  Two   Micron  All   Sky  Survey
\citep[2MASS;][]{Skrutskieetal06}   and    ALLWISE   \citep[Wide-field
  Infrared Survey Explorer;][]{Wright2018}  photometry associated with
good  quality flags,  and avoiding  upper limits  in the  fit. In  the
matching process we  took into account the \G\, proper  motions of the
targets to compute the position in the epoch 2000 and applied a search
radius of 5 arcsec.

We used  the low-mass  star CIFIST and  the hydrogen-rich  white dwarf
models  in  the  fit,   which  provided  the  bolometric  luminosities
($L_\mathrm{bol}  =  4\pi  D^{2}F_\mathrm{bol}$,   where  $D$  is  the
distance  to  each  target  and  $F_\mathrm{bol}$  is  the  bolometric
flux\footnote{See
\url{http://svo2.cab.inta-csic.es/theory/vosa/helpw4.php?otype=star&what=intro}
for a description on how  the bolometric flux is derived.}), effective
temperatures (from  the best-model fit  in the grids  after re-scaling
the  flux) and  radii  (both from  the flux  scaling  factors of  both
components   and   the   Stefan–Boltzmann  equation)   for   the   two
components. We subsequently derived  the white dwarf surface gravities
interpolating   the   effective   temperatures  and   radii   in   the
hydrogen-rich  cooling  sequences  from La  Plata  \citep{Althaus2013,
  Camisassa2016, Camisassa2019}. Finally, we  obtained the white dwarf
masses from  the well-known relation  $g=GM/R^{2}$, where $M$  and $R$
are the mass and the radius, $G$ is the gravitational constant and $g$
is  the surface  gravity  (note that  we obtained  $\log  g$ from  the
cooling sequences).

\begin{figure}
    \centering
    \includegraphics[width=0.9\columnwidth]{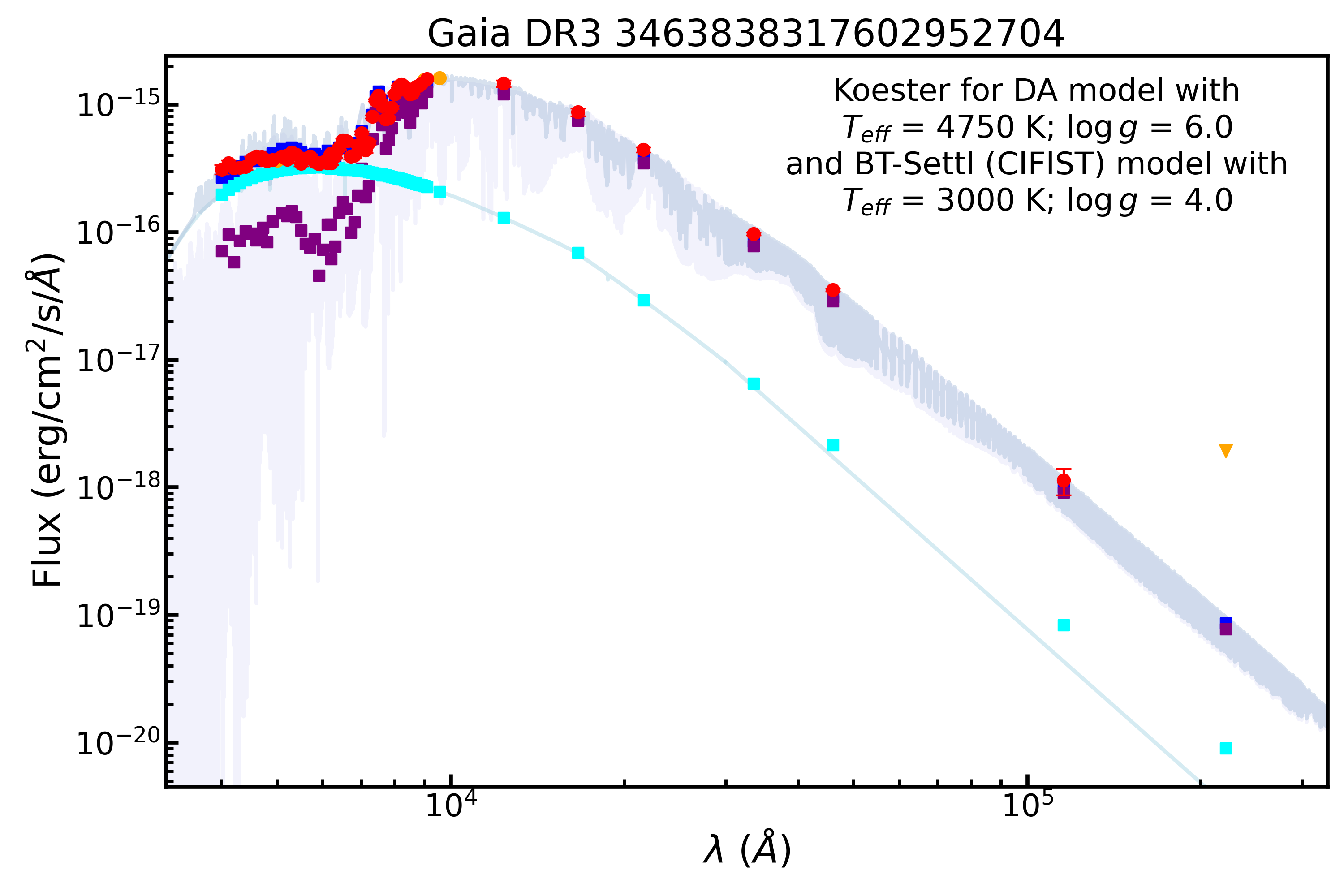} 
    \includegraphics[width=0.9\columnwidth]{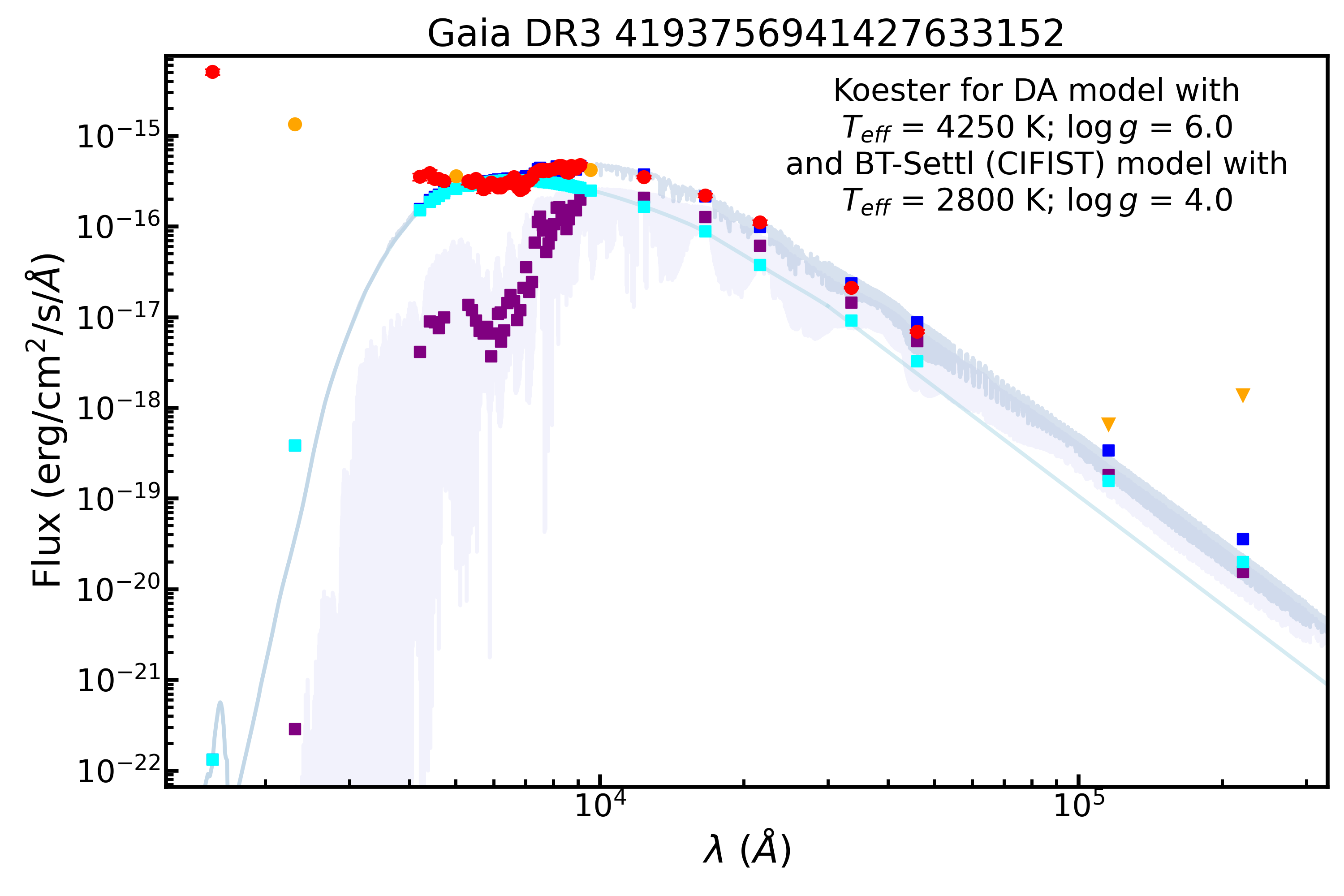} 
    \includegraphics[width=0.9\columnwidth]{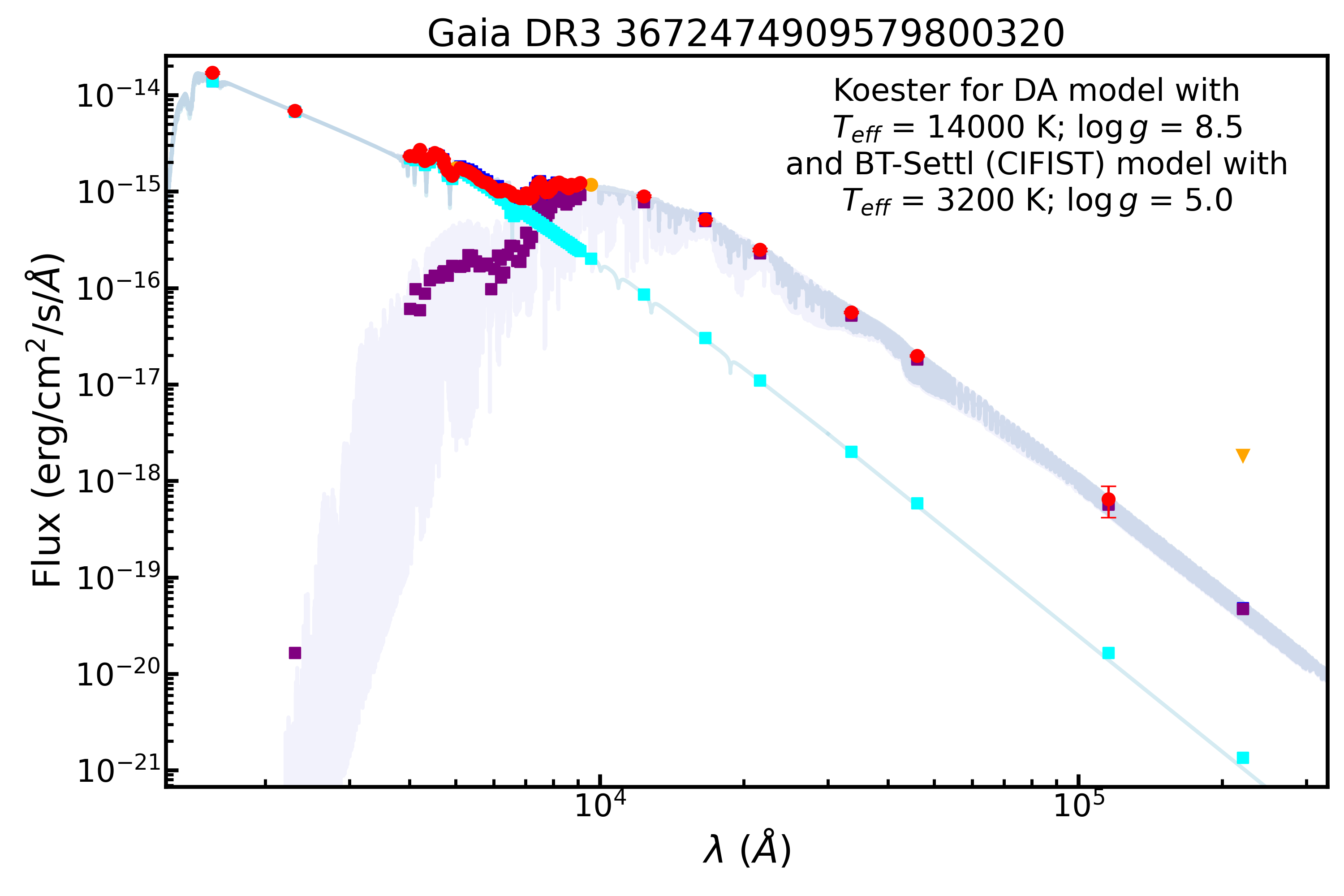} 
    \caption{Example  of two-body  VOSA fits.  The top  panel shows  a
      typical MS-dominated WDMS binary  in our sample. The combination
      of models (purple for the secondary  star and cyan for the white
      dwarf)  seem  to fit  relatively  well  the observed  data  (red
      dots). However, the effective temperature  of the white dwarf is
      too low. Clearly, such a low-luminosity white dwarf would not be
      seen  against  a 3000\,K  secondary  M  star. The  middle  panel
      illustrates  an  example  of  a   bad  fit,  especially  in  the
      ultraviolet range.  The bottom  panel shows  what we  consider a
      good  fit, where  the  models  match the  observed  data at  all
      wavelengths.}
    \label{fig_seds}
\end{figure}

We visually  inspected the two-body  fits to evaluate the  validity of
the results obtained.  It became obvious that, in the  cases where the
secondary star dominates the SED, VOSA  was generally unable to find a
combination of  models that  satisfactorily sampled the  observed SEDs
due  to  the lack  of  points  available  at  blue wavelengths.  As  a
consequence, even when the combination  of models matched the observed
data, we considered most of the  results as unreliable since the white
dwarfs  typically   piled  up   at  too  low   effective  temperatures
(5000-7000\,K)  as   compared  to  those  from   the  secondary  stars
(2800-3000\,K). Such low-luminosity white  dwarfs would not be visible
in the optical against such M  star companions. An example is shown in
the   top  panel   of   Figure\,\ref{fig_seds}.   To  compensate   the
intrinsically low  bolometric fluxes  (and luminosities) of  the white
dwarfs,  VOSA tends  to  yield  large radii  to  those objects,  which
translates into very low white  dwarf masses. This likely explains the
excess   of   extremely   low-mass    white   dwarfs   identified   in
\citet{rebassa-mansergasetal21-1}. In Brown et  al. (in prep.) we will
discuss in  detail these  issues, but  we advance  here that  all fits
resulting  in   white  dwarf  effective  temperatures   of  less  than
10\,000\,K should be  taken with caution. In other words,  the SEDs of
WDMS  with  dominating main-sequence  companions  have few  available
points at blue  wavelengths in their SEDs, where the  white dwarfs are
expected  to  contribute  most.  As a  consequence,  the  white  dwarf
parameters tend  to be  unreliable. Having available  GALEX photometry
helps to mitigate  this effect. However, the visual  inspection of the
fits also revealed that, in some cases, the best-fit white dwarf model
failed  at sampling  the GALEX  photometry  (see the  middle panel  of
Figure\,\ref{fig_seds}).

\begin{figure}
    \centering
    \includegraphics[width=0.9\columnwidth]{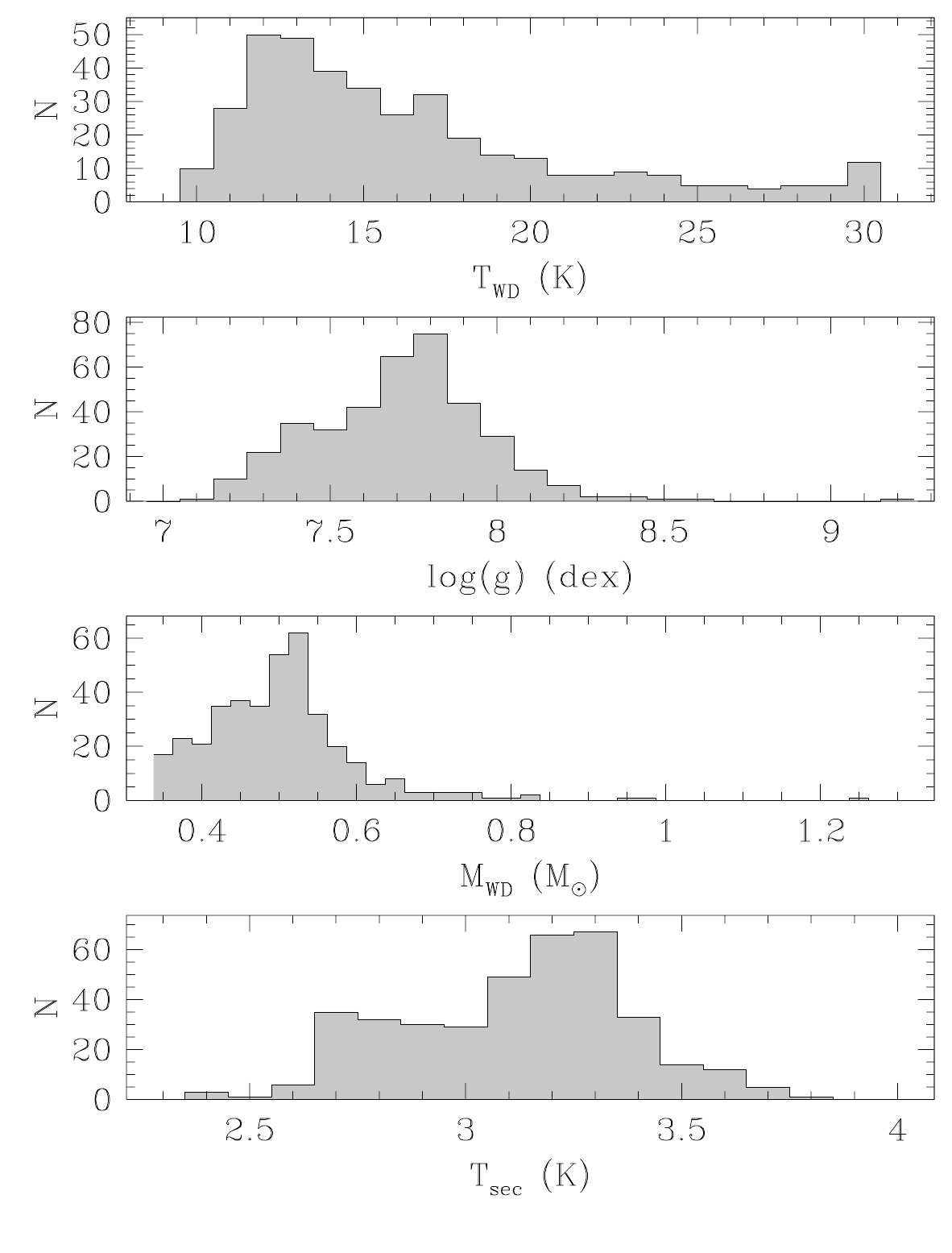}
    \caption{From  top  to bottom,  the  distribution  of white  dwarf
      effective temperatures, surface  gravities, masses and secondary
      star effective temperatures derived  from 435 WDMS binaries with
      reliable VOSA two-body fits.}
    \label{fig_param}
\end{figure}

Due to the reasons outlined above, we only considered as reliable fits
those with white dwarf  effective temperatures larger than 10\,000\,K,
white dwarf masses higher than 0.35\,$M_{\odot}$ and with the best-fit
white dwarf model matching the  GALEX data, when available. An example
is  illustrated in  the bottom  panel of  Figure\,\ref{fig_seds}. This
resulted in 435 WDMS binaries with reliable fits.  The distribution of
white dwarf effective temperatures,  surface gravities and masses, and
secondary    star     effective    temperatures    are     shown    in
Figure\,\ref{fig_param}.

Most  white dwarfs  have  effective temperatures  between 12\,000  and
17\,000\,K, with  a long  tail towards higher  temperatures, as  it is
expected   from  a   magnitude-limited  sample   (see,  for   example,
\citealt{rebassa-mansergasetal10-1}). The  white dwarf masses  peak at
$\simeq0.5\,M_{\odot}$   and   the    surface   gravities   at   $\log
g\simeq$7.8\,dex,   in  agreement   with  the   volume-limited  sample
presented  in  \citet{rebassa-mansergasetal21-1}.  In  that  paper  we
argued  these   peaks  are   lower  than   the  canonical   values  of
$\simeq0.6\,M_{\odot}$  and 8\,dex  presumably  due to  the fact  that
unresolved WDMS binaries within 100 pc are likely post common-envelope
binaries,   which    tend   to    contain   low-mass    white   dwarfs
\citep{rebassa-mansergasetal11}. As we show in Section\,\ref{s-eclip},
our sample  may indeed have  a large fraction of  post-common envelope
systems, which  is also expected  in a magnitude-limited  sample since
low-mass  white  dwarfs  are  more  luminous  for  a  fixed  effective
temperature \citep[see][for  a discussion  on how this  effect affects
  the SDSS WDMS sample]{rebassa-mansergasetal11}.  However, it is also
worth  noting   that  \citet{Santos-Garcia2025}  gave   evidence  that
$\simeq$30\%  of the  \G\,  WDMS in  \citet{rebassa-mansergasetal21-2}
passed through a common envelope  phase, a fraction that is presumable
similar, or  even smaller due  to the larger distances  considered, in
the current  sample. Therefore, the  peak at lower white  dwarf masses
might also be related to the same issue with the two-body fits in VOSA
mentioned above, which tends to yield lower masses than expected.

\begin{figure}
    \centering
    \includegraphics[width=0.9\columnwidth]{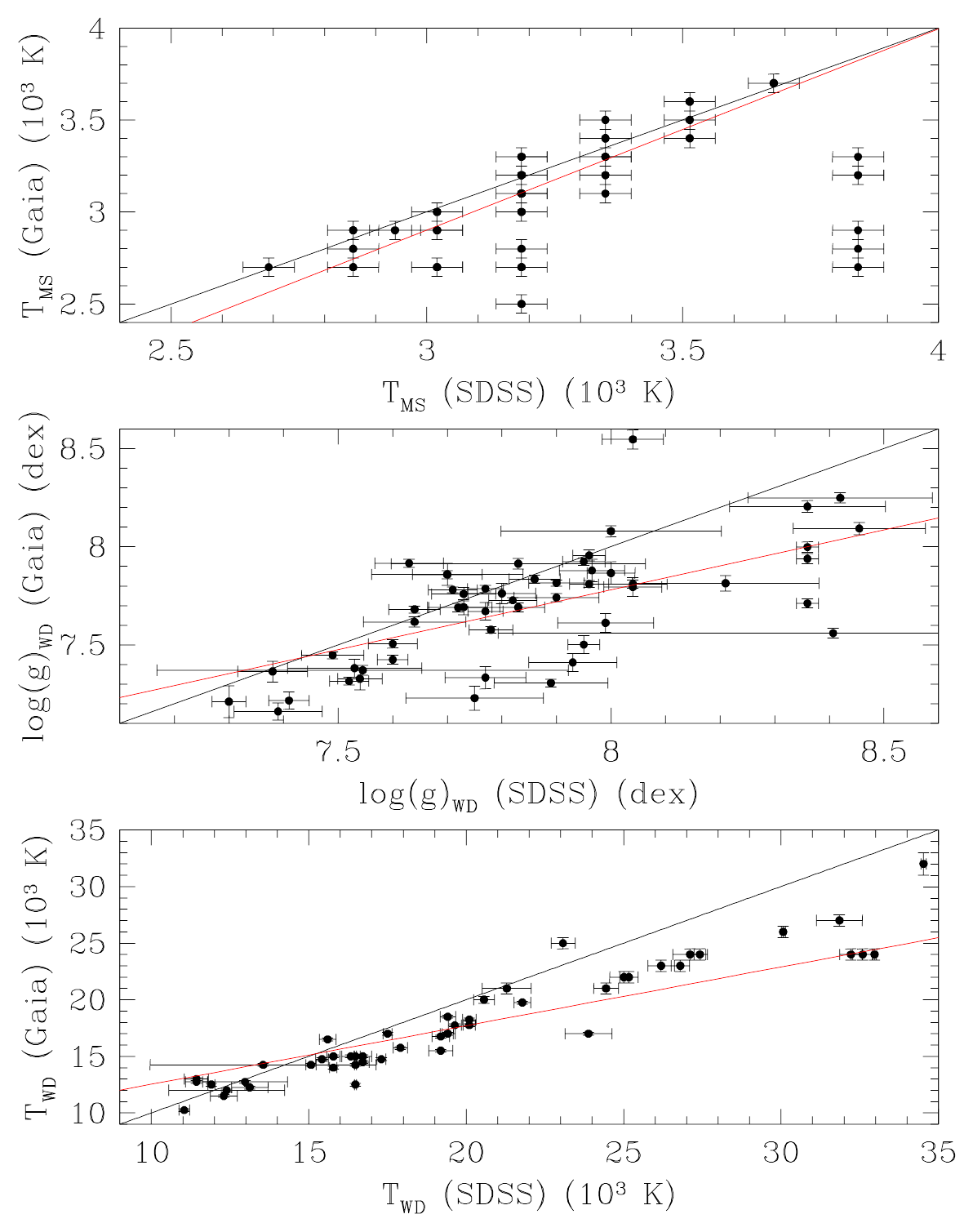}
    \caption{Comparison between the white dwarf effective temperatures
      (bottom) and surface gravities (middle) as well as main-sequence
      star effective temperatures (top) for 54 WDMS binaries with VOSA
      reliable two-body fits and spectroscopic parameters derived from
      SDSS spectra.  The black  dashed lines indicate  the one  to one
      relation, whilst  the red  dashed lines are  linear fits  to the
      data  (in  the  top  panel  the  values  above  3700\,K  in  the
      horizontal  axis  have  not  been  considered  for  being  clear
      outliers).}
    \label{fig_comp_param}
\end{figure}

In  the middle  and bottom  panels of  Figure\,\ref{fig_comp_param} we
compare the white dwarf  surface gravities and effective temperatures,
respectively, of  54 objects  with reliable VOSA  fits that  also have
such     values     derived     from    available     SDSS     spectra
\citep{rebassa-mansergasetal16-1}.  It becomes  obvious that  not only
the  surface   gravities,  hence   masses,  but  also   the  effective
temperatures derived in this work seem to be systematically lower than
those obtained fitting  the much higher resolution  SDSS spectra. This
effect may  be related to  reddening. Even though extinction  is taken
into account in  the VOSA fits, it could affect  the white dwarfs more
than their companions since they are bluer.

The  secondary star  effective  temperatures  are mainly  concentrated
between  2700  and 3400\,K,  with  a  peak  at around  3200\,K,  which
corresponds    to     M    dwarfs    of    spectral     types    M3-M6
\citep{Rebassa2007}. This is in line with our expectations, since WDMS
with earlier/later spectral type companions  would tend to fall out of
our regions of study.

The same pattern is observed when considering all WDMS binaries in the
sample,  including  those  objects   with  white  dwarfs  cooler  than
10\,000\,K  and less  massive  than 0.3\,$M_{\odot}$  as indicated  by
their VOSA white dwarf fits. We consider this as an indicator that the
secondary star  fits from VOSA  are more reliable than  those obtained
for the white  dwarfs. Indeed, when comparing  these temperatures with
those obtained  from the SDSS spectra  for the 54 common  objects with
derived   parameters\footnote{Note   that  the   decomposition/fitting
routine  of the  SDSS WDMS  binary spectra  yields secondary  spectral
types         rather        than         effective        temperatures
\citep{rebassa-mansergasetal10-1}.   We converted  the spectral  types
into temperatures using the  relation of \citet{Rebassa2007}, which is
virtually     identical      to     the     updated      tables     of
\citet{Mamajek+Hillenbrand2008}.}, we  find that  for only  12 objects
($\simeq$22\% of the  cases) the differences are of  more than 150\,K.
The  most dramatic  discrepancy  is for  five  objects with  effective
temperatures higher than  3700\,K as derived from  their SDSS spectra,
which are considered to be much  cooler according to the VOSA fits. In
these cases the temperature difference ranges from 500 to 1100\,K.

We conclude  this section by  emphasising that the  stellar parameters
obtained  from the  VOSA two-body  fits are  generally reasonable  and
reliable  for  the  secondary   stars.  Conversely,  the  white  dwarf
parameters should  only be considered  as reliable for  certain ranges
(effective temperatures larger than  10\,000\,K and masses higher than
0.3\,$M_{\odot}$),   and   especially   when   GALEX   photometry   is
available. That is, in those  cases where the secondary star dominates
the SED, too  few photometric points are available in  the blue, where
the white dwarfs mostly contribute.  As a consequence, the white dwarf
parameters  obtained   from  the   fit  are  subject   to  substantial
uncertainties. The  stellar parameters  for each target  with visually
acceptable two-body fits are included in Table\,\ref{t-full}.

\section{Comparison with other WDMS binary samples}
\label{s-comp}

In  this section  we  compare  our WDMS  binary  catalogue with  other
published  samples from  \G.. For  completeness, we  also compare  our
catalogue  to  the  largest  spectroscopic  sample  of  WDMS  binaries
previous  to  \G\,,  from  SDSS.  We  also  use  the  results  of  the
comparisons to estimate the completeness of our sample.

Note that we  do not include a  comparison with \citet{Sidharthetal24}
since their  \G\, WDMS binaries are  located in the white  dwarf locus
and are therefore excluded by our selection criteria. In the same way,
we do  not compare our catalogue  with the sample of  \G\, astrometric
binaries  from  \citet{Shahaf2024}  since  these  objects  are  mainly
located in the main-sequence, therefore  also outside of our region of
study.

\begin{figure*}
    \centering
    \includegraphics[width=0.9\columnwidth]{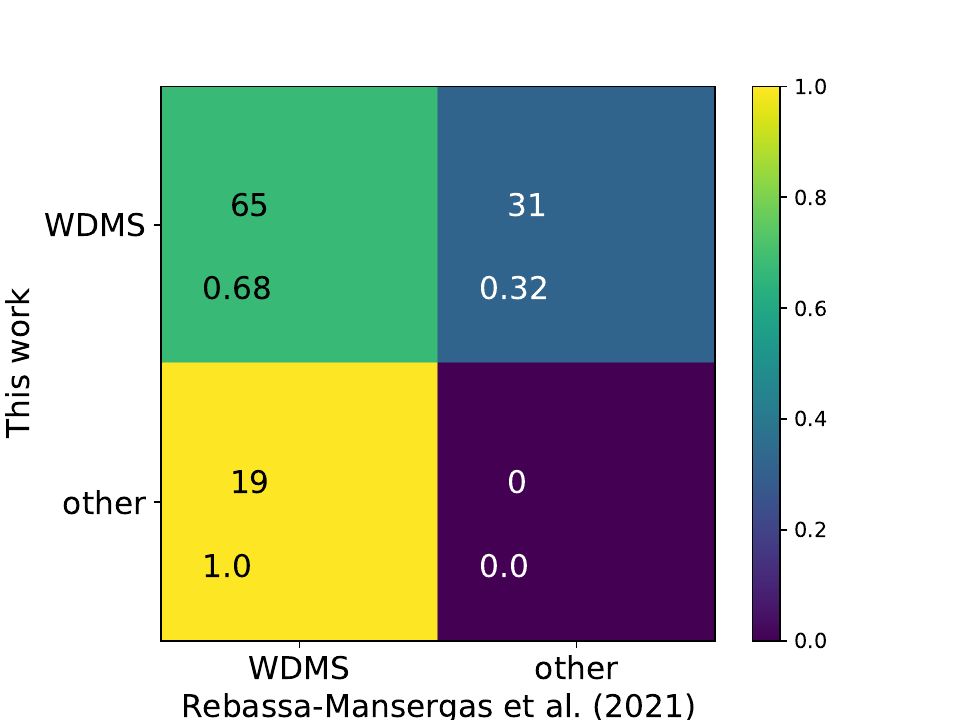}
    \includegraphics[width=0.9\columnwidth]{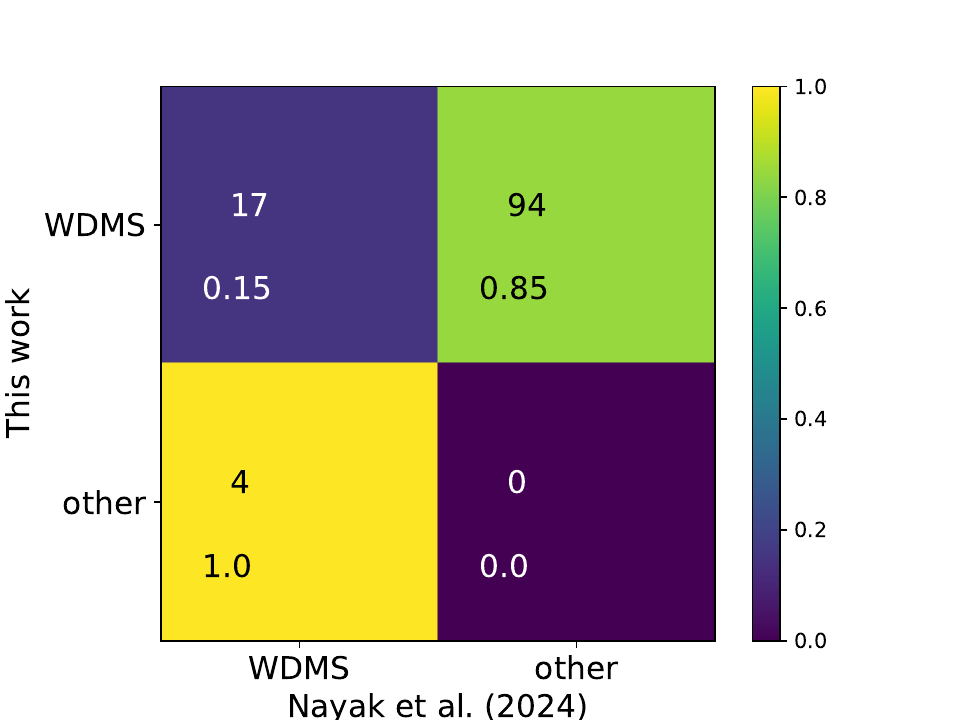}
    \includegraphics[width=0.9\columnwidth]{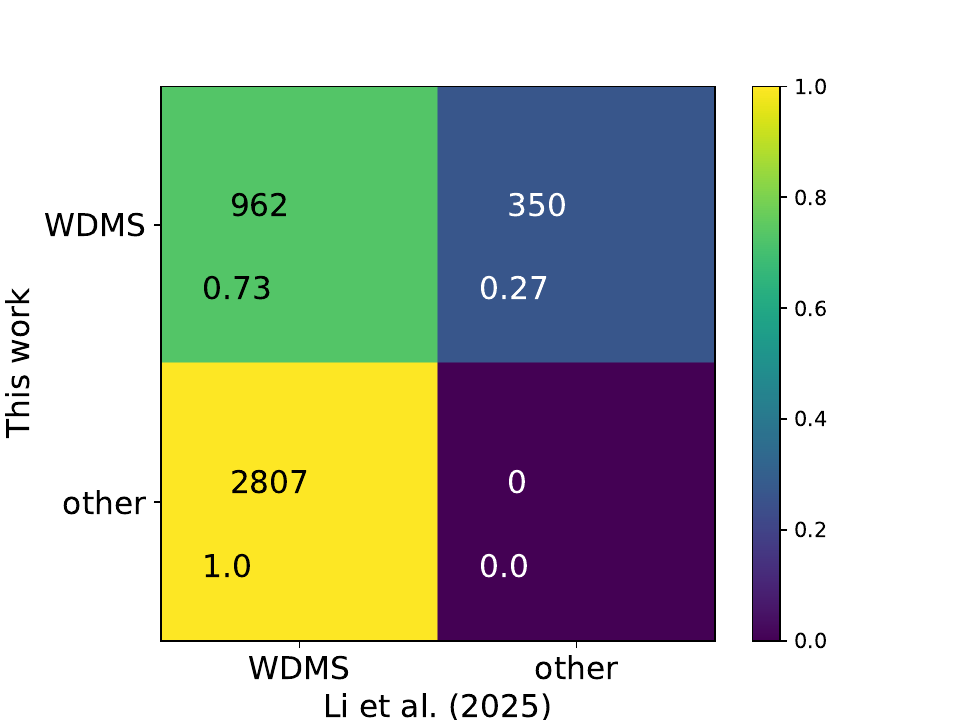}
    \includegraphics[width=0.9\columnwidth]{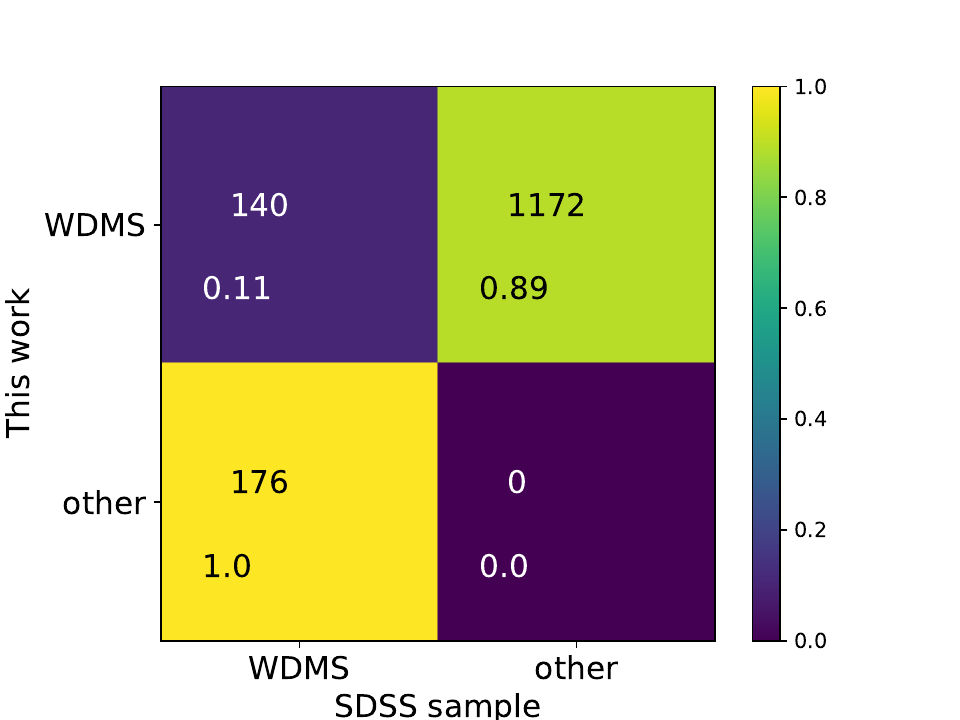}
    \caption{Confusion  matrices representing  the level  of agreement
      between      our      catalogue     and      other      samples:
      \citet{rebassa-mansergasetal21-1}           (top          left),
      \citet{Nayaketal24} (top right),  \citet{Lietal25} (bottom left)
      and the SDSS  WDMS sample (bottom right). The  values within the
      matrices indicate the number of targets falling in each category
      and the  percentages respect to  the total number of  objects in
      each  row. In  particular,  the top-left  cells  in each  matrix
      indicate the number  and fraction of common  objects, whilst the
      diagonal cells  indicate the  number of objects  (and fractions)
      missed (or not included) by the corresponding works.}
    \label{fig_cm}
\end{figure*}

\subsection{Comparison with \citet{rebassa-mansergasetal21-1}}
\label{s-100pc}

The  present  work aims  at  enlarging  the volume-limited  sample  we
provided in  \citet{rebassa-mansergasetal21-1}. Here we  check whether
or not the 112 WDMS candidates in  that sample are included in our new
catalogue. Of the  112 sources, 84 have \G\, DR\,3  spectra and 65 are
common  objects.  Of  the remaining  19 candidates,  5 and  6 are  now
classified in this work as single main-sequence stars and single white
dwarfs,  respectively,  while  performing   the  VOSA  fits  to  their
synthetic J-PAS  SEDs (Section\,\ref{s-sample}). We repeated  the fits
including GALEX, 2MASS and WISE  photometry and found the same results
except  for  two  objects:  \G\, ID  1900545847646195840  and  \G\,  ID
5490140356700680576, which display near  infrared excess arguably from
a  low-mass   companion  that   requires  confirmation.    Hence,  the
discrepancy   between  the   results  obtained   here  and   those  in
\citet{rebassa-mansergasetal21-1}  for these  objects  is  due to  the
better-sampled optical SEDs used in this work (57 points), compared to
the     considerably     fewer      optical     points     used     in
\citet{rebassa-mansergasetal21-1}.  The other 8 objects are considered
as a cataclysmic variable (1), as single white dwarfs (2) or as single
main-sequence stars (5) after visual inspection of their \G\, spectra.
It is also worth noting that our new catalogue includes 31 WDMS binary
candidates    within   100    pc   that    were   not    included   in
\citet{rebassa-mansergasetal21-1}.  In  all except 8 cases  the visual
inspection of  the 31 spectra  revealed one  of the two  components to
dominate  most of  the flux  in  the optical,  systems that  challenge
identification    by   any    method.    The    top-left   panel    of
Figure\,\ref{fig_cm}  provides  a  confusion matrix  illustrating  the
level of agreement between the catalogues.

\subsection{Comparison with \citet{Nayaketal24}}
\label{s-nayak}

\citet{Nayaketal24} provided a sample of  257 WDMS binaries within 100
pc by means  of combining optical with ultraviolet  data. This allowed
to identify WDMS binaries in the MS locus of the \G\, $G_\mathrm{abs}$
vs. $G_\mathrm{BP}-G_\mathrm{RP}$  diagram, with  28 of  their sources
falling in  our region of  study and 21 with  \G\, spectra. Of  the 21
targets, 17 are  in our list, 3  are associated to a  large excess and
one  we consider  as a  cataclysmic variable.  The top-right  panel of
Figure\,\ref{fig_cm}  provides  the   corresponding  confusion  matrix
illustrating the level of agreement between the catalogues.

\subsection{Comparison with \citet{Lietal25}}
\label{s-Lietal}

\citet{Lietal25}  has provided  a  catalogue  of $\simeq$30\,000  WDMS
binaries from \G\,  data release 3. They  used artificial intelligence
neural networks  to select them  among the millions of  available \G\,
spectra. The  advantage of their approach  is that it allowed  them to
identify systems not only in the  WDMS bridge between white dwarfs and
main-sequence  stars, but also  outside this region. Thus,  their work
introduces a new methodology for identifying WDMS that can potentially
reduce observational selection effects.

We compare our sample of 1312 WDMS with their list in this section. Of
their 30\,000 sources, 3769 are within  our WDMS binary region and 962
are common objects.  This means there are 2807  WDMS binary candidates
in their  list that we  do not have and  350 candidates from  our list
that they do not  have. Figure\,\ref{fig_cm} (bottom-left) illustrates
the corresponding confusion matrix between the catalogues.

\begin{figure}
    \centering
    \includegraphics[angle=-90,width=0.9\columnwidth]{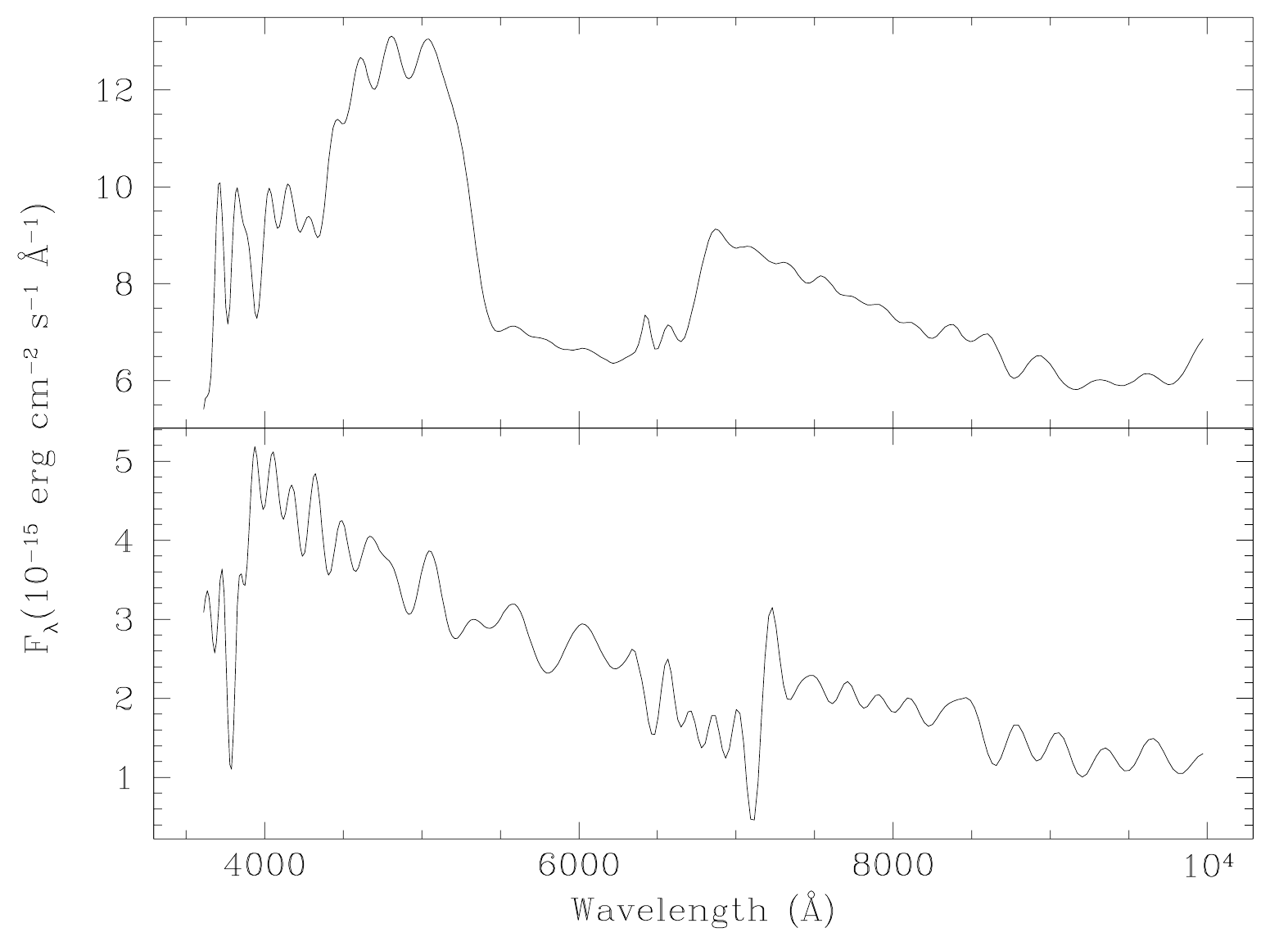}
    \caption{Example  spectra   of  WDMS  binaries  in   the  list  of
      \citet{Lietal25} that  we do not  consider as such.  \G\, source
      IDs   are  767397543537053312   (top)  and   5952567592693723904
      (bottom).}
    \label{fig_lietal}
\end{figure}

We visually inspected the 2807  objects from \citet{Lietal25} that are
not  in our  list  and  identified 111  as  WDMS.  These objects  were
selected  by our  initial parallax  and flux  cuts, but  were excluded
because of large  excess factor (82 sources)  and uncertain astrometry
(29 sources). The  remaining 2696 objects are not considered  by us as
WDMS  binaries  according to  our  visual  inspection, but  they  were
flagged as such by the artificial intelligence neural network. Indeed,
many of the spectra are  virtually identical or closely resemble those
of typical single  main-sequence stars and white dwarfs  and our human
inspection   is    unable   to   confirm   or    disprove   the   WDMS
classification. It  is therefore very  possible that several  of those
are  indeed  WDMS binaries.  However,  we  also found  35  cataclysmic
variables among the  2696 sources and many spectra  that are difficult
to interpret  as representative of  the WDMS binary population  (see a
couple of  examples in Figure\,\ref{fig_lietal}). We  consider this is
probably because \citet{Lietal25} did not visually inspect the spectra
of their 30\,000 WDMS binary  candidates to exclude objects from their
list.

The  350  objects that  are  included  in  our  catalogue but  not  in
\citet{Lietal25} all  show the  typical features  of WDMS  binaries in
their spectra, with both  components visible. \citet{Lietal25} did not
apply any cut in excess factor nor astrometric\_excess\_noise to their
sample, only on RUWE. However, the  RUWE values of the 350 sources are
generally not too large ($<\simeq$1.5) and  we therefore do not find a
clear reason why these objects were missed in their analysis.

\subsection{Comparison with the SDSS WDMS binary catalogue}
\label{s-sdss}

\begin{figure}
    \centering
    \includegraphics[width=0.9\columnwidth]{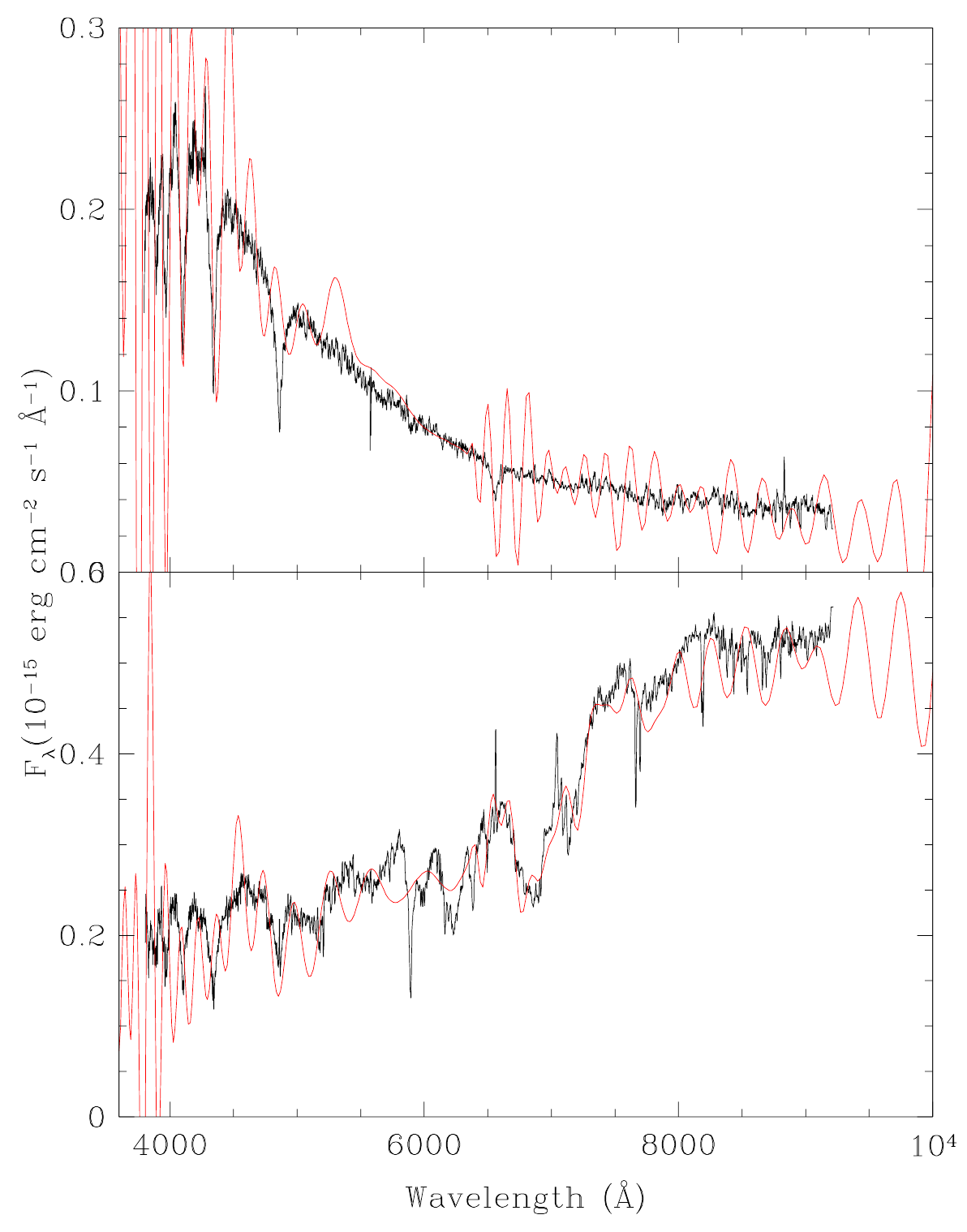}
    \caption{Example  spectra of  WDMS binaries  displaying red  (top;
      \G\,   ID  904263926328520320)   and  blue   (bottom;  \G\,   ID
      686844023151243904) excess  clearly visible in the  SDSS spectra
      (black) but diluted in the \G\, spectra (red).}
    \label{fig_excess}
\end{figure}

The spectroscopic catalogue  of WDMS binaries from  SDSS contains 3287
objects  \citep{rebassa-mansergasetal16-1},  of  which 316  have  \G\,
spectra  and  pass  our  parallax/flux relative  error  cuts  of  10\%
(Section\,\ref{s-sample}), with  140 in our  final sample. Of  the 176
that we do not have, 62 were  excluded due to a large excess factor, 4
because they had large RUWE values, 6 because they did not have enough
points for reliable VOSA fits, 86/5 because they had $\chi^{2}$ values
smaller than 10 when fitting them with VOSA using white dwarf/low-mass
star  CIFIST models  and  were  considered as  single  objects and  13
because the  visual inspection  of their \G\,  spectra did  not reveal
clear features of both components. We also visually inspected the 86+5
objects  that  we   considered  as  single  objects   based  on  their
single-body  VOSA fits  and  found  the same  issue,  i.e. their  \G\,
spectra did  not reveal  clear features  of both  objects. This  is an
observational bias  related to  the low resolution  of \G..  All these
objects reveal mild  blue or red excess in the  higher resolution SDSS
spectra that  are not featured in  the \G\, spectra. Two  examples are
shown in Figure\,\ref{fig_excess}. It is worth noting that just one of
the 91 WDMS that we characterise  as individual sources based on their
VOSA $\chi^{2}$ values is included  in the sample of \citet{Lietal25},
which  indicates  the  neural  networks also  struggle  to  find  such
objects, especially those with dominant white dwarf primaries.

In  the  bottom-right  panel   of  Figure\,\ref{fig_cm}  we  show  the
confusion  matrix  illustrating the  level  of  agreement between  our
catalogue and the SDSS sample.

\subsection{Completeness of the catalogue}

From  the  analysis  in  the previous  sections  we  identify  several
important  issues  that limit  the  completeness  of our  WDMS  binary
catalogue,    defined   as    $N_\mathrm{cat}/N_\mathrm{tot}$,   where
$N_\mathrm{cat}$ is the  number of WDMS binaries in  our catalogue and
$N_\mathrm{tot}$ is  the total  number of  observable WDMS  within the
considered      region      of     the      \G\,      $G_\mathrm{abs}$
vs.       $G_\mathrm{BP}-G_\mathrm{RP}$        diagram.       Ideally,
$N_\mathrm{cat}/N_\mathrm{tot}$ should be close to 1.

To begin  with, not  all the  WDMS binaries  with \G\,  photometry and
astrometry have spectra. For example,  of the 3287 SDSS WDMS binaries,
3089  have \G\,  photometry and  astrometry,  but only  316 have  \G\,
spectra   (Section\,\ref{s-sdss}).   In   the  100   pc   samples   of
\citet{Nayaketal24} and  \citet{rebassa-mansergasetal21-1}, 21  out of
28    and   84    out   112,    respectively,   have    \G\,   spectra
(Sections\,\ref{s-nayak}  and  \ref{s-100pc}).  Thus, at  100  pc  the
fraction of WDMS with \G\, spectra seems to be $\simeq$75\%, and drops
significantly  to  $\simeq$10\% for  distances  as  large as  1.5\,kpc
(which is  approximately the maximum  distance at which the  SDSS WDMS
binaries are located; \citealt{rebassa-mansergasetal10-1}).

In  addition,  in  Section\,\ref{s-Lietal}   we  identified  111  WDMS
binaries from \citet{Lietal25}  that are not in  our catalogue because
of   the    excess   factor    and   astrometry   cuts    applied   in
Section\,\ref{s-sample}. In the  same way, 66 SDSS  WDMS binaries with
\G\, spectra  are not in  our final list  because of the  same reasons
(Section\,\ref{s-sdss}).  Since  our sample  consists of  1312 sources
and  177 targets  are  confirmed WDMS  that  did not  make  it to  our
catalogue, this means that we missed at least $\simeq12$\% of the WDMS
binaries.   A  further  complication  is  the  fact  that  it  becomes
increasingly  more  difficult  to  identify WDMS  binaries  with  mild
blue/red excess in their optical spectra  due to the low resolution of
\G..  As mentioned in Section\,\ref{s-sdss},  91 SDSS WDMS binaries of
such characteristics were excluded from our sample since we considered
them as  individual sources based on  the results from the  VOSA fits.
Knowing that our final sample is $N_{\rm cat}=$1312 and recalling that
$N_{\rm  tot}$ is  the  total  number of  observable  WDMS within  the
considered region, then:

\begin{equation}
 N_{\rm tot}\times f_{\rm spec}\times f_{\rm cuts} \times f_{\rm vis}=N_{\rm cat},
 \label{e-ntot}
\end{equation}

\noindent where $f_{\rm  spec}$ is the fraction of  expected WDMS with
\G\, spectra  ($\simeq0.1$; this is  a lower limit since  the fraction
depends on the distance considered), $f_{\rm cuts}$ is the fraction of
expected WDMS with  \G\, spectra that we recovered  after applying our
cuts in astrometry  and excess factor ($\simeq0.88$; this  is an upper
limit since at least a fraction of  0.12 are expected to be missed, as
discussed above) and  $f_{\rm vis}$ is the fraction of  WDMS with \G\,
spectra satisfying  our quality  cuts that we  expect to  display both
components (or at least significant  blue/red excess) in their spectra
($\simeq0.6$; since 40\% of the  SDSS WDMS satisfying our quality cuts
do not show  both components in their low-resolution  \G\, spectra and
are even  considered as single  objects when performing the  VOSA fits
(Section\,\ref{s-sdss})).

From the above fractions and Equation\,\ref{e-ntot}, we derive a value
of $N_{\rm tot}=$24\,848,  that is a lower limit  for the completeness
$N_{\rm  cat}$/$N_{\rm  tot}$ of  $\simeq$5\%  for  our catalogue,  or
$\simeq$50\% if we consider WDMS with available \G\, spectra.  This is
not  surprising  given  the   large  number  of  observational  biases
involved.   On  the  positive  side, the  presented  catalogue,  which
represents the tip  of the iceberg of the  underlying WDMS population,
is statistically large  enough and is well-characterised.  That is, we
have obtained  reliable estimates of  the percentages of  WDMS systems
missed  due  to   each  observational  bias,  all  of   which  can  be
incorporated  into  numerical   simulations.   This  allows  synthetic
populations  to  be  meaningfully  compared  with  the  observed  one,
providing tighter  constraints on binary star  formation and evolution
(see,  for example,  \citealt{Santos-Garcia2025},  who performed  this
exercise  with   the  112  \G\,   WDMS  binaries  within  100   pc  of
\citealt{rebassa-mansergasetal21-1}).

\section{Eclipsing systems}
\label{s-eclip}

\begin{figure}
    \centering
    \includegraphics[width=0.9\columnwidth]{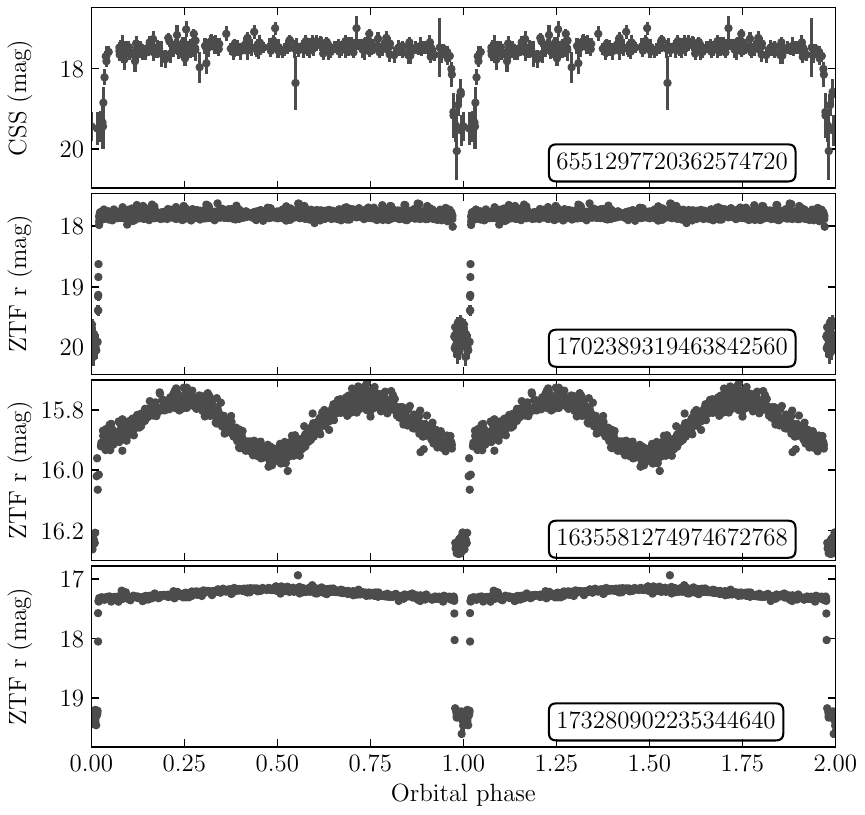}
    \caption{CRTS  and ZTF  phase-folded photometry  for 4  of the  67
      eclipsing  systems  in the  sample.  Two  orbits are  shown  for
      clarity  and the  respective  \textit{Gaia} DR3  source IDs  are
      displayed in the bottom right of each panel.}
    \label{fig_eclipsers}
\end{figure}

Given  that a  significant number  of the  systems in  our sample  are
likely  to   have  evolved   through  a  common-envelope   phase,  and
consequently have short  orbital periods, we should  expect a fraction
to have an orbital inclination such  that they are seen to eclipse. We
cross-matched the 1312 WDMS binaries  in the sample with the catalogue
of  eclipsing  WDMS  binaries   from  the  Zwicky  Transient  Facility
\citep[ZTF;][]{Dekani2020} by  van Roestel et al.  in prep.) returning
63 matches. 20 of these are in  the sub-sample of 435 systems with good
VOSA   two-body   fits  and   therefore   have   estimates  of   their
parameters. Follow-up  observations of these objects  will hence allow
determining masses  and radii that  can be directly compared  to those
estimated here. In order to identify any southern eclipsing systems we
also checked  photometry from the Catalina  Real-time Transient Survey
DR3 \citep[CRTS;][]{Drakeetal2009},  following the method  outlined in
\citet{Parsons2013}  and  only targeted  objects  outside  of the  ZTF
footprint. We found four eclipsing systems, although we note that CRTS
does not  go as deep  as ZTF. We obtained  the orbital periods  of the
eclipsing  systems we  have identified  in both  surveys via  applying
standard Lomb-Scargle periodograms to the light-curves, and then using
a  box least  squares  periodogram  to refine  the  periods (see  some
examples in Figure\,\ref{fig_eclipsers}). The  results are provided in
Table\,\ref{t-full}.

Estimates for the fraction of eclipsing post-common-envelope binaries,
PCEBs, containing a white dwarf and a low-mass main-sequence companion
typically lie  around 12-18\%  \citep{Parsons2013, Santos-Garcia2025},
but the exact  fraction depends on the orbital  period distribution as
well  as the  mass, and  therefore  radius, distributions  of the  two
stars. The number of eclipsing systems within the sample can therefore
provide a  lower limit on  the fraction of  PCEBs in our  catalogue (a
lower limit  because not all  eclipsing systems will be  detected). Of
the 1312  systems in the full  sample, 920 of these  are accessible to
the ZTF survey  (declination > -31 degrees), of which  63 are found to
be eclipsing. Assuming  that eclipsing systems account  for 12-18\% of
PCEBs, this suggests that at  least $\simeq38-57$\% of the full sample
are  PCEBs. Performing  the same  analysis  for the  sub-sample of  435
systems    with   reliable    fits    to    the   \G\,    spectroscopy
(Section\,\ref{s-param} and  Figure\,\ref{fig_param}) we find  that at
least $\simeq31-46$\% of these are PCEBs.

The  estimated PCEB  fraction among  WDMS in  our sample  seems to  be
higher  than  expected.   Numerical simulations  indicate  that  PCEBs
account  for  approximately  25-30\%  of  the  total  WDMS  population
\citep{Willems+Kolb04, Toonen+Nelemans13}, including  the \G\, 100\,pc
sample \citep{Santos-Garcia2025}. Observational studies reveal similar
PCEB        fractions        \citep{Schreiber2010,        Nebotetal11,
  rebassa-mansergasetal11}.   It  is  therefore  plausible  that  some
observational biases  affect the  \G\, WDMS  sample, which  favour the
detection  of  eclipsing  systems.    For  example,  as  mentioned  in
Section\,\ref{s-param},  a magnitude-limited  WDMS sample  favours the
detection of low-mass white dwarfs,  since these are more luminous for
a   fixed   effective   temperature   \citep{rebassa-mansergasetal11}.
Moreover, the wider the orbital separations, the more likely it is for
\G\, to resolve  the two components, which would imply  a bias towards
shorter  period systems.   In the  same  way, wide  binaries are  more
likely to be associated to larger values of RUWE and/or excess factor,
objects that  may be excluded by  our quality cuts.  It  is also worth
noting  that  our  estimated  PCEB  fractions  assume  that  the  cuts
performed in magnitude-colour space do  not impact the likelihood of a
PCEBs to eclipse.  For a more  accurate estimate of the PCEB fraction,
population  synthesis  techniques (e.g.   \citealt{Santos-Garcia2025})
will be required.

\section{Summary and conclusions}
\label{s-concl}

During the last  two decades it has been shown  that WDMS binaries are
of great use to improve our understanding of a wide range of topics in
astronomy. This  relies on  the existence  of statistically  large and
well-defined samples  that allow  characterising the  biases affecting
the observed populations. In this sense,  in this work we have built a
sample of 1312  WDMS binaries via mining the  spectroscopic content of
the data release 3 of \G\.

The  catalogue is  expected to  be $\simeq$50  per cent  complete. The
missing targets  are predominantly expected  to be objects  with large
values of RUWE and/or the excess  noise parameters, as well as objects
with mild blue  or red excess in their optical  spectra, features that
are diluted in the low-resolution \G\, spectra.  The identification of
such  WDMS is  expected to  improve by  using artificial  intelligence
algorithms  applied   to  the  \G\,   spectra  \citep{Echeverryetal22,
  Lietal25,  Perez-Couto2025},   although  they   fail  to   detect  a
non-negligible  fraction  of  WDMS   binaries  and  often  misclassify
irregular spectra.   Moreover, the completeness dramatically  drops to
$\simeq$5 per  cent (lower  limit) if  we consider  that not  all WDMS
binaries  in  \G\, have  available  spectra.   However, despite  these
issues,  our  catalogue  is  very   well  characterised  in  terms  of
implemented photometric and astrometric  cuts and observational biases
and hence can be directly compared to synthetic samples that reproduce
the WDMS  binary population in  the Galaxy to constrain,  for example,
binary  star  evolution  \citep{Santos-Garcia2025}. In  addition,  the
study  of exotic  objects in  the sample,  such as  the 67  identified
eclipsing systems (which represent 5\% of the total sample), allows to
place tighter  constraints on the  mass-radius relation of  both white
dwarfs   and   low-mass    main-sequence   stars   \citep{Parsons2017,
  Parsons2018}.

We  find the  catalogue  to  be dominated  by  binaries  in which  the
main-sequence  companion  contributes  more in  the  optical  spectral
energy distribution. Because  of this, the stellar  parameters that we
derived  for most  of  the white  dwarfs in  these  objects should  be
considered with caution.   Future follow-up spectroscopic observations
at higher resolution are therefore desired for better characterization
of the white dwarfs. Hence,  these sources are excellent supplementary
targets   for    the   forthcoming    White   Dwarf    Binary   Survey
\citep{Toloza2023} implemented  in the  overall 4MOST  survey (4-metre
Multi-Object Spectroscopic Telescope; \citealt{deJong2022}).

\section{Data availability}

Table\ref{t-full} is only available in  electronic form at the CDS via
anonymous   ftp   to   cdsarc.u-strasbg.fr   (130.79.128.5)   or   via
http://cdsweb.u-strasbg.fr/cgi-bin/qcat?J/A+A/.

\begin{acknowledgements}

We thank Enrique Garc\'ia-Zamora for  helping with the construction of
the  confusion  matrices.  We  thank the  anonymous  referee  for  the
comments and suggestions on the manuscript.

This  work  was  partially  supported  by  the  Spanish  MINECO  grant
PID2023-148661NB-I00 and  by the AGAUR/Generalitat de  Catalunya grant
SGR-386/2021. This work was partially supported by the Spanish Virtual
Observatory  (\url{https://svo.cab.inta-csic.es)   project  funded  by
 MCIN/AEI/10.13039/501100011033/}             through            grant
PID2023-146210NB-I00.   RMO   is   funded  by   INTA   through   grant
PRE-OBSERVATORIO. RR acknowledges support from Grant RYC2021-030837-I,
funded  by  MCIN/AEI/  10.13039/501100011033 and  by  “European  Union
NextGeneration  EU/PRTR”.  MC   acknowledges  grant  RYC2021-032721-I,
funded  by MCIN/AEI/10.13039/501100011033  and by  the European  Union
NextGenerationEU/PRTR.  This  work presents results from  the European
Space Agency (ESA) space mission \G..  \G, data are being processed by
the \G\, Data  Processing and Analysis Consortium  (DPAC). Funding for
the  DPAC is  provided  by national  institutions,  in particular  the
institutions   participating  in   the  \G\,   MultiLateral  Agreement
(MLA).        The        \G\,         mission        website        is
\url{https://www.cosmos.esa.int/gaia}.  The  \G\, archive  website  is
\url{https://archives.esac.esa.int/gaia}. This job has made use of the
Python package  GaiaXPy, developed  and maintained  by members  of the
\G\,  Data   Processing  and   Analysis  Consortium  (DPAC),   and  in
particular, Coordination Unit 5 (CU5),  and the Data Processing Centre
located at the Institute of Astronomy, Cambridge, UK (DPCI).

\end{acknowledgements}

\end{document}